\documentclass[iop, tighten, twocolappendix]{emulateapj13}

\usepackage{pslatex}
\usepackage{epstopdf}
\usepackage{amsmath}
\usepackage{textcomp}

\usepackage{hyperref}
\hypersetup {
    bookmarks=false,                    
    pdftitle={Magnetic Inhibition of Convection and the Fundamental Parameters of Low-Mass Stars. I. Stars with a Radiative Core},                   
    pdfauthor={Gregory A. Feiden},     
    pdfsubject={},                     
    colorlinks=true,                   
    linkcolor=blue,                    
    citecolor=blue,                    
    urlcolor=blue                      
}
\newcommand{\myemail}{gregory.a.feiden.gr@dartmouth.edu}
\newcommand{\bcemail}{brian.chaboyer@dartmouth.edu}

\newcommand{\thermd}[4] {\left(\frac{\partial #1}{\partial #2}\right)_{#3,\, #4}}
\newcommand{\el}[2]{$^{#1}$#2}               
\newcommand{\msun}{M_{\odot}}                
\newcommand{\rsun}{R_{\odot}}                
\newcommand{\lsun}{L_{\odot}}                
\newcommand{\teff}{T_{\rm eff}}              
\newcommand{\porb}{P_{\rm orb}}              
\newcommand{\amlt}{\alpha_{\rm MLT}}         
\newcommand{\dela}{\nabla_{\rm ad}}          
\newcommand{\dele}{\nabla_{\rm e}}           
\newcommand{\delx}{\nabla_{\rm \chi}}        
\newcommand{\tgrad}{\nabla_{\rm s}}          
\newcommand{\uconv}{u_{\rm conv}}            
\newcommand{\deltamm}{\delta_{\rm MM}}       

\shortauthors{Feiden \& Chaboyer}
\slugcomment{Accepted for Publication in the Astrophysical Journal}
\journalinfo{The Astrophysical Journal, 779:1 (25pp) 2013 ???}
\submitted{Received: 2013 August 20; accepted 2013 October 31; published 
           2013 ???}

\begin{document}

\title{Magnetic Inhibition of Convection and the Fundamental Properties \\
       of Low-Mass Stars. I. Stars with a Radiative Core}

\author{Gregory A. Feiden\altaffilmark{1} and Brian Chaboyer}
\affil{Department of Physics and Astronomy, Dartmouth College, 6127 Wilder 
Laboratory, Hanover, NH 03755, USA; \\ \href{mailto:\myemail}{\myemail}, 
\href{mailto:\bcemail}{\bcemail}}

\altaffiltext{1}{Current address: Department of Physics and Astronomy, Uppsala University,
Box 516, SE-751 20 Uppsala, Sweden.}

\begin{abstract}
Magnetic fields are hypothesized to inflate the radii of low-mass stars---defined
as less massive than $0.8\msun$---in detached eclipsing binaries (DEBs).
We investigate this hypothesis using the recently introduced magnetic 
Dartmouth stellar evolution code. In particular, we focus on stars thought 
to have a radiative core and convective outer envelope by studying in detail
three individual DEBs: UV Psc, YY Gem, and CU Cnc. Our results suggest that
the stabilization of thermal convection by a magnetic field is a plausible 
explanation for the observed model-radius discrepancies. However, surface 
magnetic field strengths required by the models are significantly stronger 
than those estimated from observed coronal X-ray emission. Agreement
between model predicted surface magnetic field strengths and those inferred
from X-ray observations can be found by assuming that the magnetic field
sources its energy from convection. This approach makes the transport of
heat by convection less efficient and is akin to reduced convective mixing length 
methods used in other studies. Predictions for the metallicity and magnetic
field strengths of the aforementioned systems are reported. We also develop
an expression relating a reduction in the convective mixing length to a
magnetic field strength in units of the equipartition value. Our results 
are compared with those from previous investigations to incorporate magnetic 
fields to explain the low-mass DEB radius inflation. Finally, we explore 
how the effects of magnetic fields might affect mass determinations using 
asteroseismic data and the implication of magnetic fields on exoplanet studies. 
\end{abstract}

\section{Introduction}
\label{sec:intro}

Magnetic fields are a ubiquitous feature of stars across the 
Hertzsprung-Russell diagram. Despite their ubiquity, magnetic fields
have often been excluded from low-mass stellar evolutionary calculations
as there has been little need for their inclusion. Recently, however,
observations of low-mass stars---here defined to have $M<0.8\msun$---in
detached eclipsing binaries (DEBs) have altered this perception; magnetic
fields might be necessary after all \citep{Ribas2006,Lopezm2007}. We began
an effort to address this necessity in a previous paper, where we described
a new stellar evolution code that includes effects due to magnetic 
perturbations \citep{FC12b}. A single case study provided an initial assessment
of the code's viability, but did not specifically investigate the problems
with low-mass stars. Here, we investigate the hypothesis that magnetic field 
effects are required to accurately model low-mass stars.

The geometry of DEBs permits a nearly model independent determination of
the fundamental properties (mass, radius, effective temperature) of the 
constituent stars \citep[see the reviews by][]{Popper1980,Andersen1991,
Torres2010}. 
Stellar masses and radii can typically be determined with a precision below 
3\% given quality photometric and spectroscopic observations \citep{Andersen1991,
Torres2010}.\footnote{One must be mindful that larger 
systematic uncertainties may be lurking in the data \citep{Morales2010,
Windmiller2010}.} This permits rigorous tests of stellar evolution theory.
Any disagreements between observations and stellar evolution models become
strikingly apparent.

Observations show that stellar evolution models routinely predict radii 
about 5\% smaller than real stars, at a given mass \citep[see, e.g.,][]{
Torres2002,Ribas2006,Morales2008,Morales2009a,Torres2010,
Kraus2011,Irwin2011,Doyle2011,Winn2011}. Star-to-star
age and metallicity variations may account for some, but not all, of the
noted discrepancies \citep{FC12,Torres2013}. However, the problem appears 
to be exacerbated by more well-studied systems, which exhibit near-10\% 
radius discrepancies \citep{Popper1997,FC12,Terrien2012}. To further complicate 
the matter, seemingly hyper-inflated stars show radii inflated by more 
than 50\% \citep{Vida2009a,Cakirli2010,Ribeiro2011,Cakirli2013a,Cakirli2013b}.
Whether the mechanism puffing up the hyper-inflated stars is related to 
the more common sub-10\% inflation is unclear. Regardless, numerous low-mass
stars show significant departures from radius predictions of standard stellar 
evolution theory.

In addition to the radius discrepancies, effective temperatures ($\teff$s)
predicted by stellar evolution models are inconsistent with observations.
Observations indicate that low-mass stars tend to be 3\% -- 5\% cooler 
than theoretical predictions \citep{Torres2010}. However, this problem is
complicated by the fact that absolute $\teff$ measurements for stars 
in DEBs are subject to significant uncertainty \citep{Torres2010,Torres2013}. 
DEB geometry only allows for an accurate determination of the temperature ratio.
Difficulty in determining absolute $\teff$s has garnered support from 
a noted discrepancy in the radius--$\teff$ relation between single field 
stars and stars in DEBs \citep{Boyajian2012}. Whether this discrepancy is 
indicative of an innate difference between single field stars and stars
in DEBs, or highlights errors in the determination of $\teff$s in either 
population is debatable. 
As a result, mass-$\teff$ discrepancies have not received as much 
attention in the literature as the mass-radius problem. We will continue 
this trend and use the mass-radius relation as a primary guide for testing 
stellar models. DEB $\teff$s will be consulted only for additional guidance. 

Other areas of astrophysics are impacted by the inaccuracies of 
stellar evolution models.
With typical lifetimes greater than a Hubble time \citep{Laughlin1997}, 
low-mass stars are excellent objects for studying galactic structure and 
evolution \citep[e.g.,][]{Reid1995,Fuchs2009,Pineda2013}. The history of
the galaxy is effectively encoded within the population of low-mass stars.
Understanding their properties allows for this history to be constructed.
Their low-mass, small radius, and faint luminosity also provides an advantage 
for observers searching Earth-sized planets orbiting in their host
star's habitable zone \citep{Charbonneau2009,Gillon2010}. Significant 
effort is being devoted to hunting for and characterizing planets 
around M-dwarfs \citep[e.g.,][]{NC08,Quirrenbach2010,Muirhead2012,Mahadevan2012,
Dressing2013}. These applications require an intimate 
understanding of how physical observables of M-dwarfs are influenced by 
the star's fundamental properties and vice-versa. It is therefore prudent 
to look closely at the problems presented by stars in DEBs to better comprehend 
the impact of a star's physical ``ingredients'' on its observable 
properties. 

The leading hypothesis to explain the model-observation radius discrepancies
is the presence of magnetic fields \citep{MM01,Ribas2006,Lopezm2007,Morales2008}. 
Many stars that display inflated radii exist in short period
DEBs whose orbital periods are less than 3\,days. Stars in short period DEBs
will have a rotational period synchronized to their orbital period by tidal 
interactions with their companion \citep{Zahn1977}. At a given main-sequence (MS)
age, stars in short period DEBs will be rotating faster than a comparable 
single field star. Since the stellar dynamo mechanism is largely driven 
by rotation, tidal synchronization allows a star to produce and maintain
a strong magnetic field throughout its MS lifetime. 

However, radius deviations are not only observed among stars in short period 
systems. A number of long period DEBs have low-mass stars that display
inflated radii \citep{Irwin2011,Doyle2011,Winn2011,Bender2012,Kep38,Welsh2012}.
Stars in long period systems, even if they are rotationally synchronized, 
are presumed to be slow rotators. In two of these systems, \object{LSPM J1112+7626} 
\citep{Irwin2011} and \object{Kepler-16} \citep{Winn2011}, this assumption has been
confirmed. The primary star in LSPM J1112+7626 has an approximately 65 day
rotation period inferred from out-of-eclipse light curve modulation, suggesting
it is both slow rotating and not rotationally synchronized. Kepler-16 A,
on the other hand, was observed to have a rotation period of roughly 36
days from spectroscopic line broadening with minimal chromospheric activity
apparent from Ca {\sc ii} observations \citep{Winn2011}.

Though these systems appear to refute the magnetic field hypothesis, little
is known about the rotational characteristics of the secondary stars. If 
the stars are spinning down as single stars \citep{Skumanich1972}, then
it is possible that the secondary stars are still rotating rapidly enough
to drive a strong dynamo. Low-mass stars appear to only require rotational
velocities of order 3\,km\,s$^{-1}$ (rotation period of roughly 3 days) 
before they display evidence of magnetic flux saturation \citep{Reiners2009a}.
Furthermore, pseudo-synchronization may take place \citep{Hut1981}.
Numerous short tidal interactions at periastron can cause binary companions
to synchronize with a period not quite equal to the orbital period \citep[see
e.g.,][]{Winn2011}. Thus, stars in long period DEBs do not necessarily 
evolve as if they were isolated, potentially exciting the stellar dynamo.
However, each of the above circumstances do not appear sufficient to 
explain the inflated radii of LSPM J1112+7626 B and Kepler-16 B. LSPM J1112
would be nearly 9\,Gyr old given the rotation period of the primary, suggesting
the secondary has also had ample time to spin down. Kepler-16 shows 
evidence for pseudo-synchronization, which would impart a rotation period
of nearly 36 days onto the secondary, giving it a rotational velocity of 
below 0.5\,km\,s$^{-1}$.

Support is lent to the magnetic field hypothesis by observations that 
low-mass DEBs exhibit strong magnetic activity. Inflated stars, in particular,
often display strong chromospheric H$\alpha$ emission \citep{Morales2008,Stassun2012} 
and strong coronal X-ray emission \citep{Lopezm2007,FC12}. Both are thought
to be indicative of magnetic fields heating the stellar atmosphere. Magnetic
activity levels may also correlate with radius inflation \citep{Lopezm2007,
Stassun2012}, but it is still an open question \citep{FC12}. Such a
correlation would be strong evidence implicating magnetic fields as the
culprit of radius inflation.

Indirect measures of magnetic field strengths (i.e., magnetic activity 
indicators) yield tantalizing clues about the origin of the observed radius 
inflation, but direct measurements are far more preferable. Although no 
direct observations of surface magnetic fields on low-mass DEBs have 
been published,\footnote{\citet{Morin13cs17} report the observations, but 
not the results, of such an endeavor.} there has been a concerted 
effort to measure surface magnetic field strengths of single low-mass stars
\citep[e.g.,][]{Saar1996,Reiners2007,JK2007,Reiners2009,Morin2010,Shulyak2011,
Reiners2012a}. K- and M-dwarfs have been a focus of magnetic field studies 
because around mid-M spectral type, about $0.35\msun$, M-dwarfs become 
fully convective \citep{Limber1958a,BCAH98}. Standard descriptions
of the stellar dynamo mechanism posit that magnetic fields are generated 
near the base of the outer convection zone \citep{Parker1955,Parker1979}. 
A strong shear layer, called the tachocline, forms between the differentially 
rotating convection zone and the radiation zone, which rotates as a solid 
body. Fully convective stars, by definition, do not possess a tachocline.
Thus, according to the standard Parker dynamo model, this would leave  
fully convective stars unable to generate or sustain a strong magnetic 
field.

\renewcommand{\arraystretch}{1.1}
\begin{deluxetable*}{l c c c c c c}[t]
    \tablewidth{2\columnwidth}
    \tablecaption{Sample of DEBs whose Stars Possess a Radiative Core.}
    \tablehead{
        \colhead{DEB}      & \colhead{Star}     &  \colhead{$\porb$}  &  
        \colhead{Mass}     & \colhead{Radius}   &  \colhead{$\teff$}  &
        \colhead{$[{\rm Fe/H}]$} \\
        \colhead{System}   & \colhead{}         &  \colhead{(day)}    &  
        \colhead{($\msun$)}& \colhead{($\rsun$)}&  \colhead{(K)}      &
        \colhead{(dex)}
    }
    \startdata
        UV Psc & A &  0.86  & 0.9829  $\pm$ 0.0077   & 1.110   $\pm$ 0.023  & 5780 $\pm$ 100 & \nodata \\
        UV Psc & B &        & 0.76440 $\pm$ 0.00450  & 0.8350  $\pm$ 0.0180 & 4750 $\pm\ $  80 & \nodata \\
        YY Gem & A &  0.81  & 0.59920 $\pm$ 0.00470  & 0.6194  $\pm$ 0.0057 & 3820 $\pm$ 100 &  +0.1 $\pm$ 0.2 \\
        YY Gem & B &        & 0.59920 $\pm$ 0.00470  & 0.6194  $\pm$ 0.0057 & 3820 $\pm$ 100 &  +0.1 $\pm$ 0.2 \\
        CU Cnc & A &  2.77  & 0.43490 $\pm$ 0.00120  & 0.4323  $\pm$ 0.0055 & 3160 $\pm$ 150 & \nodata \\
        CU Cnc & B &        & 0.39922 $\pm$ 0.00089  & 0.3916  $\pm$ 0.0094 & 3125 $\pm$ 150 & \nodata
    \enddata
    \label{tab:deb_rad_core}
\end{deluxetable*}

Despite lacking a tachocline, low-mass stars are observed to
possess strong magnetic fields with surface strengths upward of a 
few kilogauss \citep{Saar1996,Reiners2007,Reiners2009a,Reiners2010a,Shulyak2011}. 
Instead of a dynamo primarily powered by rotation, turbulent convection 
may be driving the stellar dynamo \citep{Durney1993,Dobler2006,Chabrier2006}.
Large-scale magnetic field topologies of low-mass stars appear to shift
from primarily non-axisymmetric to axisymmetric across the fully convective 
boundary \citep{Morin2008,Donati2009,PhanBao2009,Morin2010}. This apparent
shift in field topology is suggested as the hallmark of a transitioning 
dynamo. However, shifts in field topology are still a subject for debate
\citep{Donati2009,Reiners2012a}. It is also uncertain whether the transition
from a rotational to a turbulent dynamo occurs abruptly at the fully convective 
boundary or if it is a gradual transition developing between early- 
and mid-M-dwarfs.

With the dynamo dichotomy in mind, we have elected to divide our analysis 
of the low-mass stellar mass-radius problem into two parts. The first 
part, presented in this paper, concerns low-mass stars in DEBs that
should possess a radiative core and convective outer envelope. The second
part, pertaining to fully convective low-mass stars, is presented in
a companion paper (G.~A. Feiden \& B. Chaboyer, in preparation). Our motivation for 
splitting the analysis is that models described in \citet{FC12b} 
assume that energy for the magnetic field---and thus the dynamo 
mechanism---is supplied by rotation. This was explicitly stated following
the discussion of Equation (41) in that paper. With the onset of complete 
convection near $M=0.35\msun$, a transition from a rotationally driven 
interface dynamo to a turbulent dynamo must occur. Thus, the theory that 
we present in \citet{FC12b} is probably not suitable for models of 
fully convective stars. Whether our theory is valid for partially convective
stars is an answer that will be pursued in this work.

We present results from detailed modeling of three DEB systems with partially
convective stars. We study only three systems to avoid muddling the results
while still providing a rigorous examination of the models. The three DEBs 
selected for analysis were \object[UV Psc]{UV Piscium} \citep{Carr1967,Popper1997}, 
\object[YY Gem]{YY Geminorum} \citep{Adams1920,Torres2002}, and 
\object[CU Cnc]{CU Cancri} \citep{Delfosse1999,
Ribas2003}. We recall their properties in Table \ref{tab:deb_rad_core}. 

These three particular
systems were chosen for three reasons: (1) they are well-studied, meaning they
have precisely determined masses and radii with reasonable estimates of 
their effective temperatures,
(2) they show large discrepancies with models \citep{FC12}, and
(3) they span an interesting range in mass, covering nearly the full range
of masses for partially convective low-mass stars.
This latter fact will allow us to assess the validity of our modeling
assumptions as we approach the fully convective boundary. Effectively, we 
will probe whether an interface dynamo of the type presented by \citet{Parker1955} 
is sufficient to drive the observed inflation, or if a turbulent dynamo 
is required to deplete the kinetic energy available in convective flows.

The paper is organized as follows: a discussion of the adopted stellar 
models is presented in Section \ref{sec:models}. In 
Section \ref{sec:ind_deb}, we demonstrate that our models are able to 
reconcile the observed radius and $\teff$ discrepancies. Discussion 
presented in Section \ref{sec:mag_strengths}, however, leads us to believe
that magnetic field strengths required by our models are unrealistic.
We therefore explore various means of reducing the surface magnetic 
field strengths. A further discussion of our results is given in Section
\ref{sec:disc}. We provide comparisons of different models and to previous 
studies. Implications for asteroseismology studies and exoplanet investigations
are also considered. Finally, we summarize the key conclusions in 
Section \ref{sec:summ}.

\section{Dartmouth Magnetic Stellar Evolution Code}
\label{sec:models}

Stellar evolution models were computed as a part of the Dartmouth Magnetic
Evolutionary Stellar Tracks and Relations program \citep{FC12b,
FeidenTh}. The stellar evolution code is a modified version 
of the existing Dartmouth stellar evolution code \citep{Dotter2008}. 
Physics used by the standard (i.e., non-magnetic) Dartmouth code have 
been described extensively in the literature \citep[e.g.,][]{Dotter2007,
Dotter2008,Feiden2011,FC12,FC12b} and will not be reviewed here. We
note that we have updated the nuclear reaction cross sections to those
recommended by \citet{FusionII}. The latest recommendations include
a revised cross section for the primary channel of the proton--proton ({\it p-p})
chain, but it does not significantly impact low-mass stellar evolution.

Effects of a globally pervasive magnetic field are included following
the prescription described by \citet{FC12b}, which is heavily based on the procedure
outlined by \citet{LS95}. Perturbations to the canonical stellar structure 
equations are treated self-consistently by considering thermodynamic 
consequences of stresses associated with a static magnetic 
field. Modifications to the standard convective mixing length
theory \citep[MLT; e.g.,][]{bv58} are derived self-consistently by assuming the
magnetic field is in thermodynamic equilibrium with the surrounding plasma. 
All transient magnetic phenomena that act to remove mass, such as 
flares and coronal mass ejections, are ignored. We also neglect the steady 
removal of mass through magnetized stellar winds. There does not 
appear to be significant mass loss from low-mass stars \citep[and references
therein]{Laughlin1997}. 

Input variables for stellar evolution models are defined 
relative to calibrated solar values. These input variables include the
stellar mass, the initial mass fractions of helium and heavy elements 
($Y_{i}$ and $Z_{i}$, respectively), and the convective mixing length 
parameter, $\amlt$. The latter defines the length scale of a turbulent 
convective eddy in units of pressure scale heights. 
Since they are all defined relative to the solar values, we must
first define what constitutes the Sun for the model setup. To do this,
we require a $1.0\msun$ model to reproduce the solar radius, luminosity,
radius to the base of the convection zone, and the solar photospheric 
$(Z/X)$ at the solar age \citep[4.57 Gyr;][]{Bahcall2005}. Adopting the
solar heavy element abundance of \citet{GS98}, our models require 
$Y_{\rm init} = 0.27491$, $Z_{\rm init} = 0.01884$, and $\amlt = 1.938$. 
The final solar model properties are given in Table \ref{tab:solar_calib}.

\renewcommand{\arraystretch}{1.1}
\begin{deluxetable}{l c c c}[h]
    \tablewidth{\columnwidth}
    \tablecaption{Solar Calibration Properties} 
    \tablehead{
        \colhead{Property}  &  \colhead{Adopted}  &  \colhead{Model}  &
        \colhead{Reference}
    }
    \startdata
    Age (Gyr) & $4.57$ & $\cdots$ & 1 \\
    $\msun$ (g) & $1.9891\times 10^{33}$ & $\cdots$ & 2  \\
    $\rsun$ (cm) & $6.9598\times 10^{10}$ & $\log(R/\rsun) = 8\times 10^{-5}$ & 3, 1 \\
    $\lsun$ (erg s$^{-1}$) & $3.8418\times 10^{33}$ & $\log(L/\lsun) = 2\times 10^{-4}$ & 1 \\
    $R_{\rm bcz}/\rsun$ & $0.713\pm0.001$ & $0.714$ & 4, 5 \\
    $(Z/X)_{\rm surf}$ & $0.0231$ & $0.0230$ & 6 \\
    $Y_{\odot,\, \rm surf}$ & $0.2485\pm0.0034$ & $0.2455$ & 7
    \enddata
    \tablerefs{(1)~\citet{Bahcall2005}; 
    (2)~IAU 2009 
    (3)~\citet{Neckel1995}; (4)~\citet{Basu1997};~(5) \citet{Basu1998};
    (6)~\citet{GS98}; (7)~\citet{Basu2004}.}
    \label{tab:solar_calib}
\end{deluxetable}

\section{Analysis of Individual DEB Systems}
\label{sec:ind_deb}

\subsection{UV Piscium}
\label{sec:uvpsc}

UV Piscium (HD 7700; hereafter UV Psc) contains a solar-type 
primary with a mid-K-dwarf companion. Numerous determinations 
the fundamental stellar properties have been performed since its discovery, 
with the most precise measurements produced by \citet{Popper1997}. These 
measurements were later slightly revised by \citet{Torres2010}, 
who standardized reduction and parameter extraction routines for a 
host of DEB systems. The mass and radius for each component of UV Psc 
recommended by \citet{Torres2010} is given in Table \ref{tab:deb_rad_core}.
No metallicity estimate exists, despite the system being relatively bright 
($V = 9.01$) and having a nearly total secondary eclipse.

One notable feature of UV Psc is that the secondary
component is unable to be properly fit by standard stellar evolution models
at the same age as the primary \citep[see e.g.,][]{Popper1997,
Lastennet2003,Torres2010,FC12}. The secondary's radius 
appears to be approximately 10\% larger than models predict and the 
effective temperature is about 6\% cooler than predicted. Metallicity
and age are known to affect the stellar properties predicted by models, 
typically allowing for better agreement with observations. However, even
when allowing for age and metallicity variation, the best fit models of
UV Psc display large disagreements \citep{FC12}. 

An investigation by \citet{Lastennet2003} found that it was possible to 
fit the components on the same theoretical isochrone. Their method 
involved independently 
adjusting the helium mass fraction $Y$, the metal abundance $Z$, and the 
convective mixing length $\amlt$. The authors were able to 
constrain a range of $Y$, $Z$, and $\amlt$ values that produced 
stellar models compatible with the fundamental properties of each component 
while enforcing that the stars be coeval. \citet{Lastennet2003} found 
that a sub-solar metal abundance ($Z = 0.012$)\footnote{ We calculate this 
implies [Fe/H] $= -0.14$ considering the required $Y$ and the fact that 
they were using the \citet{GN93} heavy element abundances.} combined with 
an enhanced helium abundance ($Y$ = 0.31) and drastically reduced mixing 
lengths for each star produced the best fit at an age of 1.9 Gyr. The final
mixing lengths were $\amlt = 0.58\alpha_{\rm MLT,\, \odot}$ and 
$0.40\alpha_{\rm MLT,\, \odot}$, for the primary and secondary, respectively, 
where $\alpha_{\rm MLT,\, \odot}$ is the solar calibrated mixing length.
The age inferred from their models is a factor of four lower than the 8 Gyr 
age commonly cited for the system. 

Despite properly fitting the two components, the investigation by \citet{Lastennet2003}
did not provide any physical justification for the reduction in mixing length.
Furthermore, they required an abnormally high helium abundance given the required
sub-solar heavy element abundance. Assuming that $Y$ varies linearly with $Z$
according to the formula
\begin{equation}
    Y = Y_{p} + \left(\frac{\Delta Y}{\Delta Z}\right)\, \left( Z - Z_{p}\right),
    \label{eq:helium}
\end{equation}
where $Y_{p}$ is the primordial helium mass fraction and $Z_p$ is the primordial
heavy element abundance ($Z_{p} = 0$), implies that $\Delta Y / \Delta Z > 5$ 
for the \citet{Lastennet2003} study. Empirically determined values typically 
converge around $2 \pm 1$ \citep{Casagrande2007}. The empirical relation
is by no means certain and there is no guarantee that all stars conform 
to this prescription. However, a single data point suggesting 
$\Delta Y / \Delta Z > 5$ is a significant outlier, at $3\sigma$ above the 
empirical relation. This introduces some doubt as to whether that 
particular $Y$ and $Z$ combination is realistic. Though we cannot definitively
rule out the results of the \citet{Lastennet2003} study, we seek an alternative
explanation to reconcile the stellar models with observations of the secondary.

The stars in UV Psc exhibit strong magnetic activity, 
showcasing a wide variety of phenomena. Soft X-ray emission \citep{Agrawal1980}, 
Ca {\sc ii} H \& K emission \citep{Popper1976, Montes1995a}, and H$\alpha$ emission 
\citep{Barden1985,Montes1995c} have all been observed
and associated with UV Psc. Flares have been recorded in H$\alpha$ 
\citep{Liu1996} and at X-ray wavelengths \citep{Caillault1982}, further suggesting
the components are strongly active. Star spots betray their presence 
in the modulation and asymmetries of several light curves \citep{UVPsc2005}.
Although some of these modulations have also been attributed to intrinsic 
variability in one of the components \citep{Antonopoulou1987}, there does
not appear to be any further evidence supporting this claim \citep{Ibanoglu1987,
Popper1997}. This leads us to believe any observed light curve variations 
are the result of spots.

\begin{figure*}[t]
    \plottwo{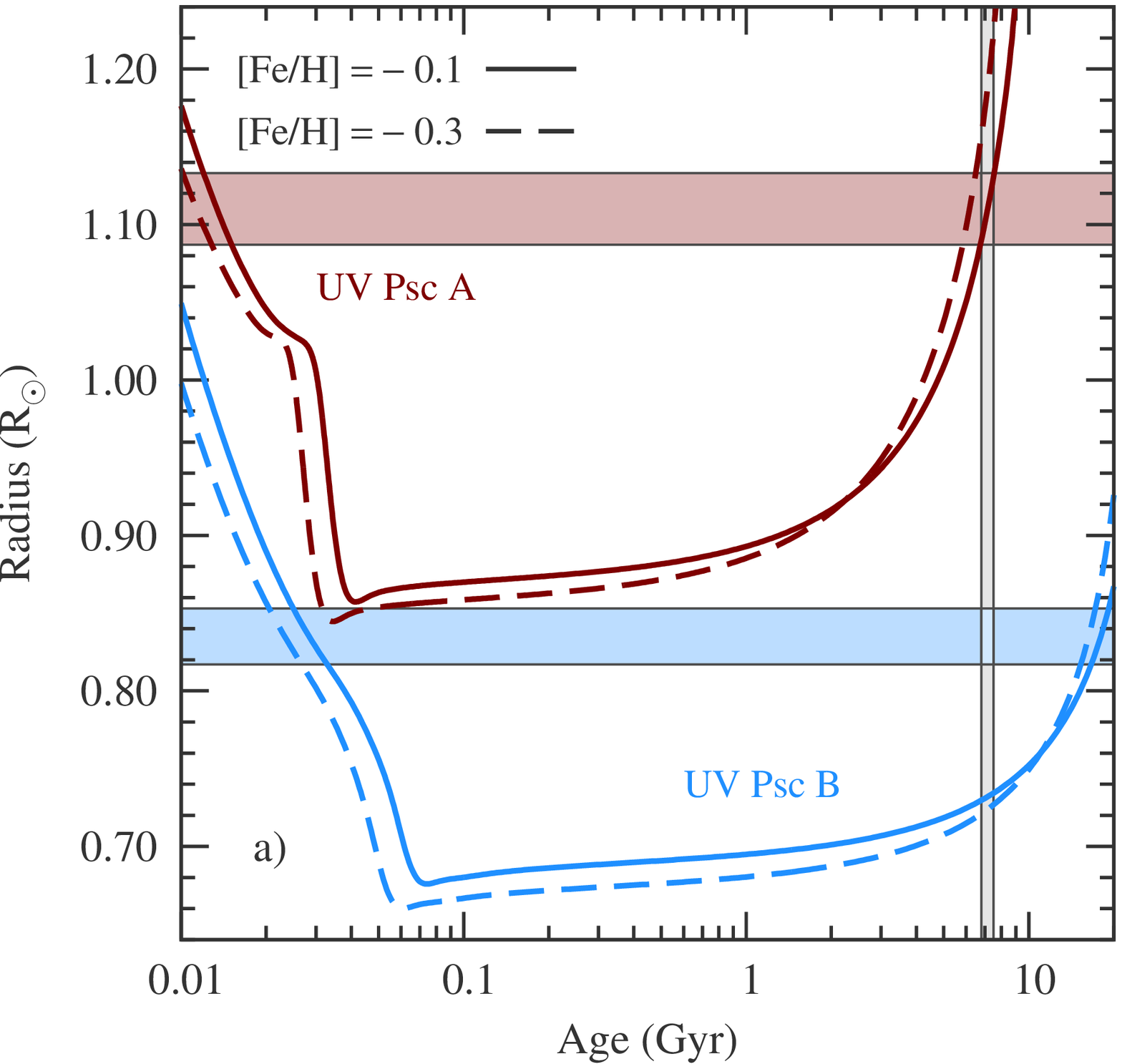}{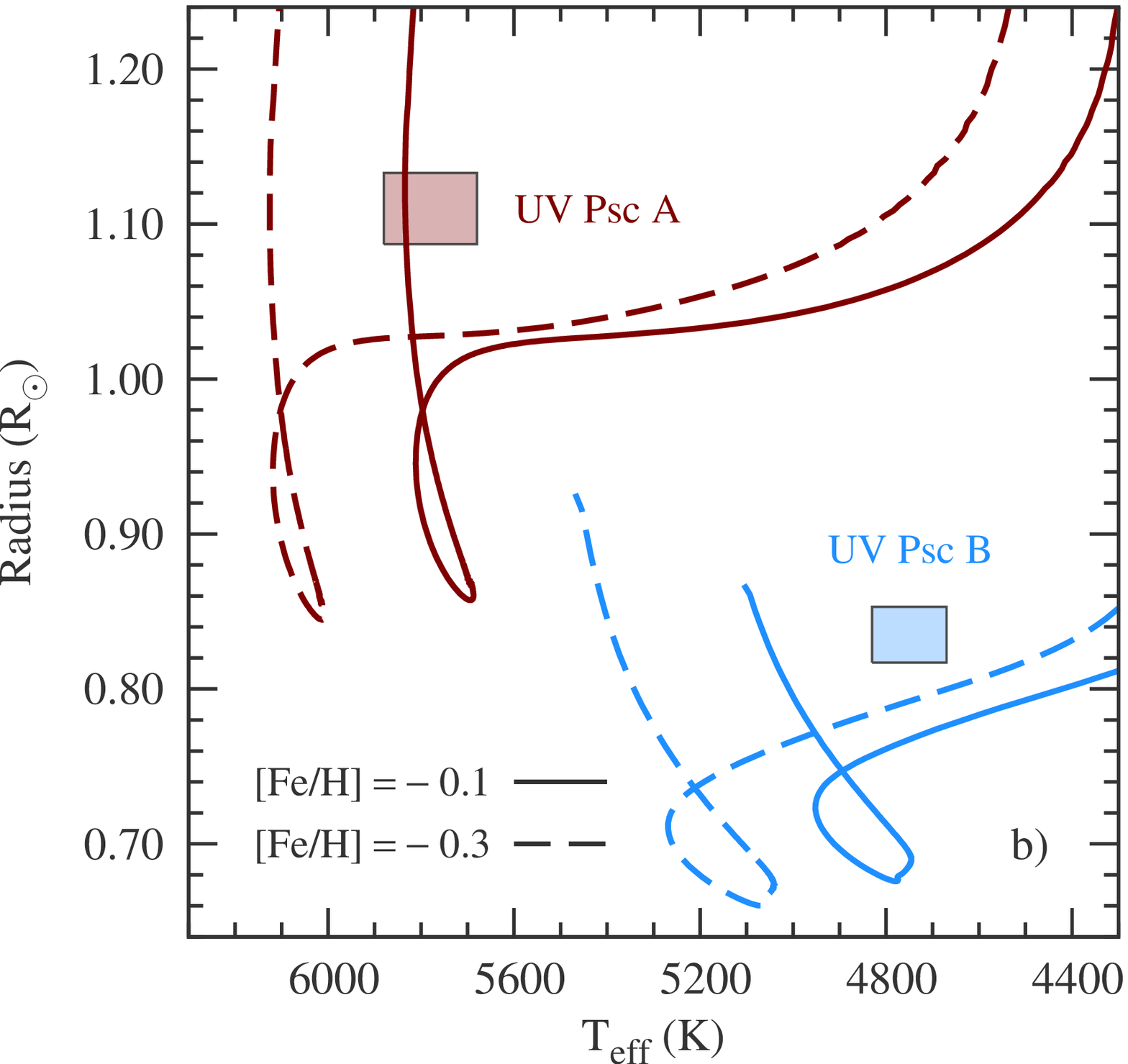}
    \caption{Standard Dartmouth mass tracks of UV Psc A (maroon)
        and UV Psc B (light blue) computed with [Fe/H] $= -0.1$ (solid 
        line) and [Fe/H] $= -0.3$ (dashed line). (a) The age-radius 
        plane. Horizontal swaths denote the observed radii with 
        associated $1\sigma$ uncertainty. The vertical region indicates
        the age predicted by the primary. (b) The $\teff$-radius
        plane. Shaded regions denote the observational constraints.
        \vspace{0.5\baselineskip}\\
        (A color version of this figure is available in the online journal.)}
    \label{fig:uvpsc_nmag}
\end{figure*}

\begin{figure*}[t]
    \plottwo{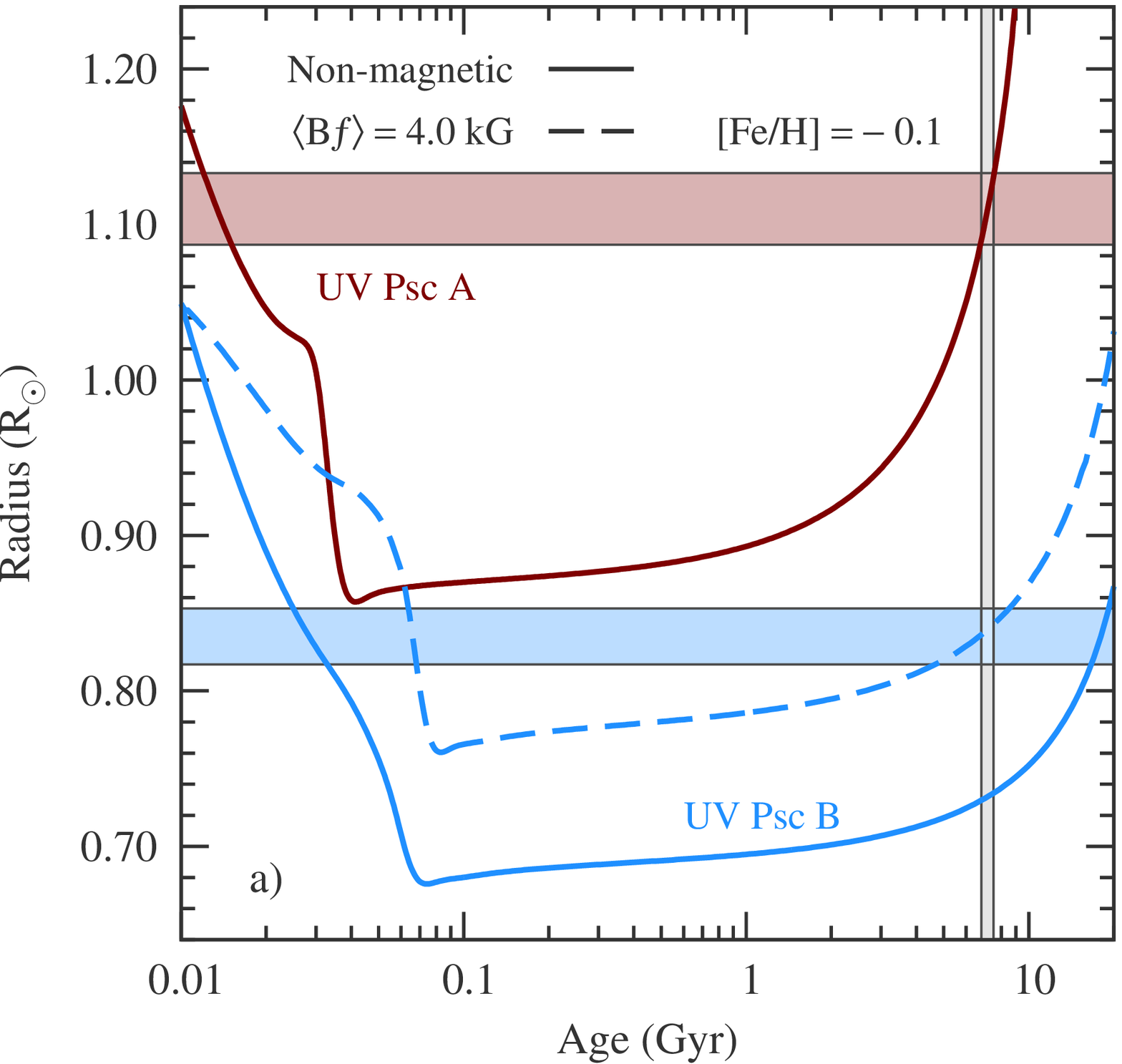}{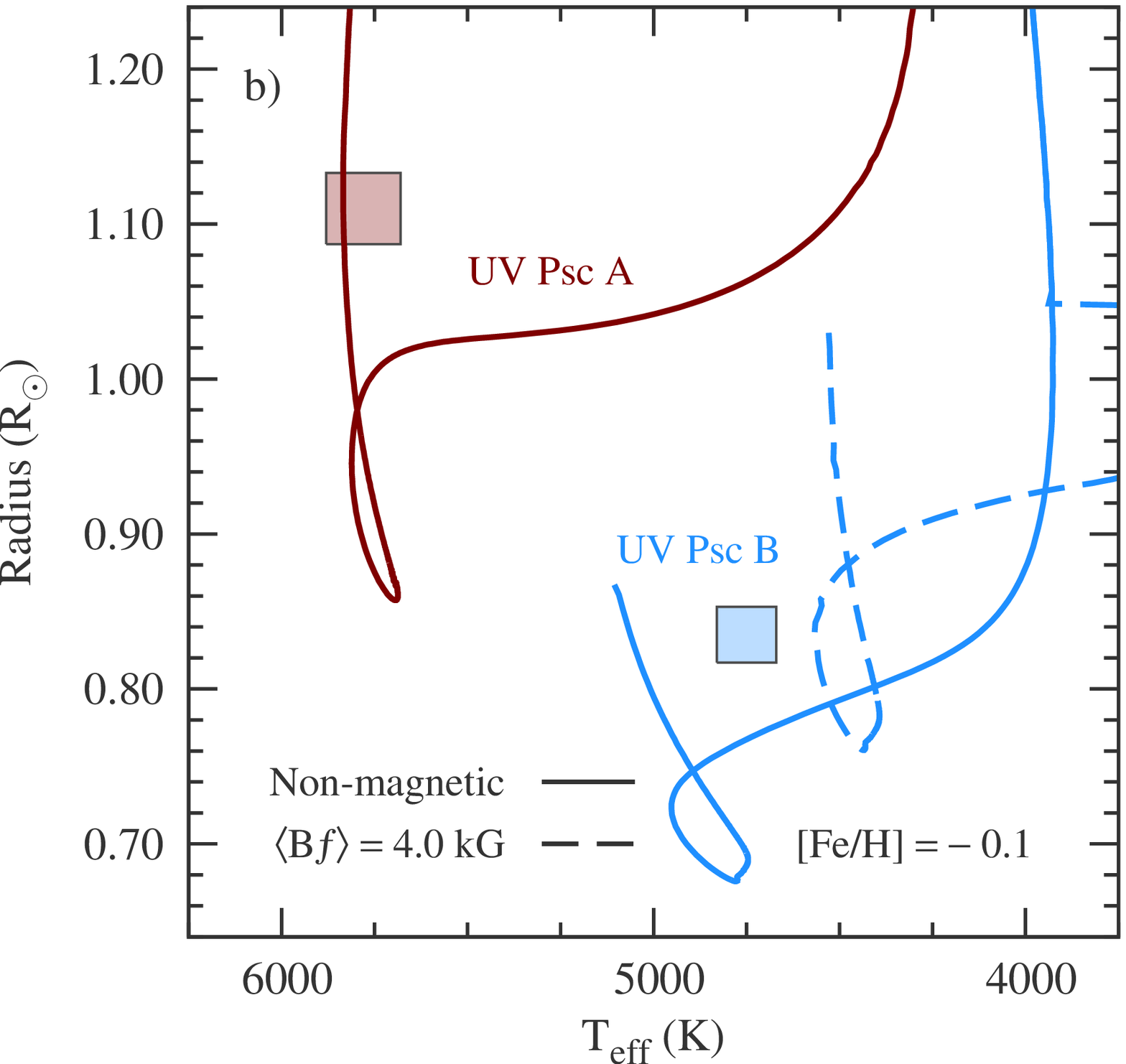}
    \caption{Similar to Figure \ref{fig:uvpsc_nmag} but with
        a single metallicity of [Fe/H] $= -0.1$ dex. Magnetic mass track
        for UV Psc B with a $4.0$ kG surface magnetic field strength 
        (light blue, dashed line). Standard Dartmouth mass tracks are plotted 
        for comparison. (a) Age-radius plane. (b) $\teff$-radius plane.
        \vspace{0.5\baselineskip}\\
        (A color version of this figure is available in the online journal.)}
    \label{fig:uvpsc_feh_b}
\end{figure*}

The aforementioned evidence provides clues that magnetic fields may be 
the source of the observed radius discrepancies. \citet{Lastennet2003}'s 
finding that a reduced convective mixing length was required could then
be explained by magnetic inhibition of thermal convection \citep{
Cox1981,Chabrier2007}.

Previous studies of UV Psc have found that standard stellar evolution models 
are able to reproduce the fundamental stellar properties of the primary
star \citep{Popper1997,Lastennet2003,Torres2010,FC12}. Therefore,
we begin by assuming that UV Psc A conforms to the predictions of stellar
evolution theory, but that magnetic effects must be invoked to reconcile
models with UV Psc B. Given this assumption, UV Psc A may be used to 
constrain the age and metallicity of the system. Using a large grid of 
stellar evolution isochrones, \citet{FC12} found UV Psc A was best fit 
by a 7 Gyr isochrone with a slightly metal-poor composition of $-0.1$ dex. 
The metallicity estimate is consistent with \citet{Lastennet2003}, though 
two independent methods were utilized to achieve the result. We adopt 
this sub-solar value as the initial target age and metallicity for the 
system.

Standard model mass tracks are illustrated in Figures~\ref{fig:uvpsc_nmag}(a) 
and (b) for two different metallicities. The age of 
the system is anchored to the narrow region in Figure~\ref{fig:uvpsc_nmag}(a) 
where the models agree with the observed primary radius. 
Figure~\ref{fig:uvpsc_nmag}(b) indicates that the [Fe/H] $= -0.1$ model 
yields satisfactory agreement with the observed radius and effective temperature.
We infer an age of 7.2 Gyr, which is more precise than \citet{FC12} 
as we are not constrained to a discretized set of isochrone ages.  
Standard models for the secondary are shown to reach the observed radius at 
an age of 18 Gyr, according to Figure~\ref{fig:uvpsc_nmag}(a). This implies
an 11 Gyr difference between the two components. We also see that the model 
effective temperature of the secondary is too hot compared to observations
by about 250 K.

\begin{figure*}[t]
    \plottwo{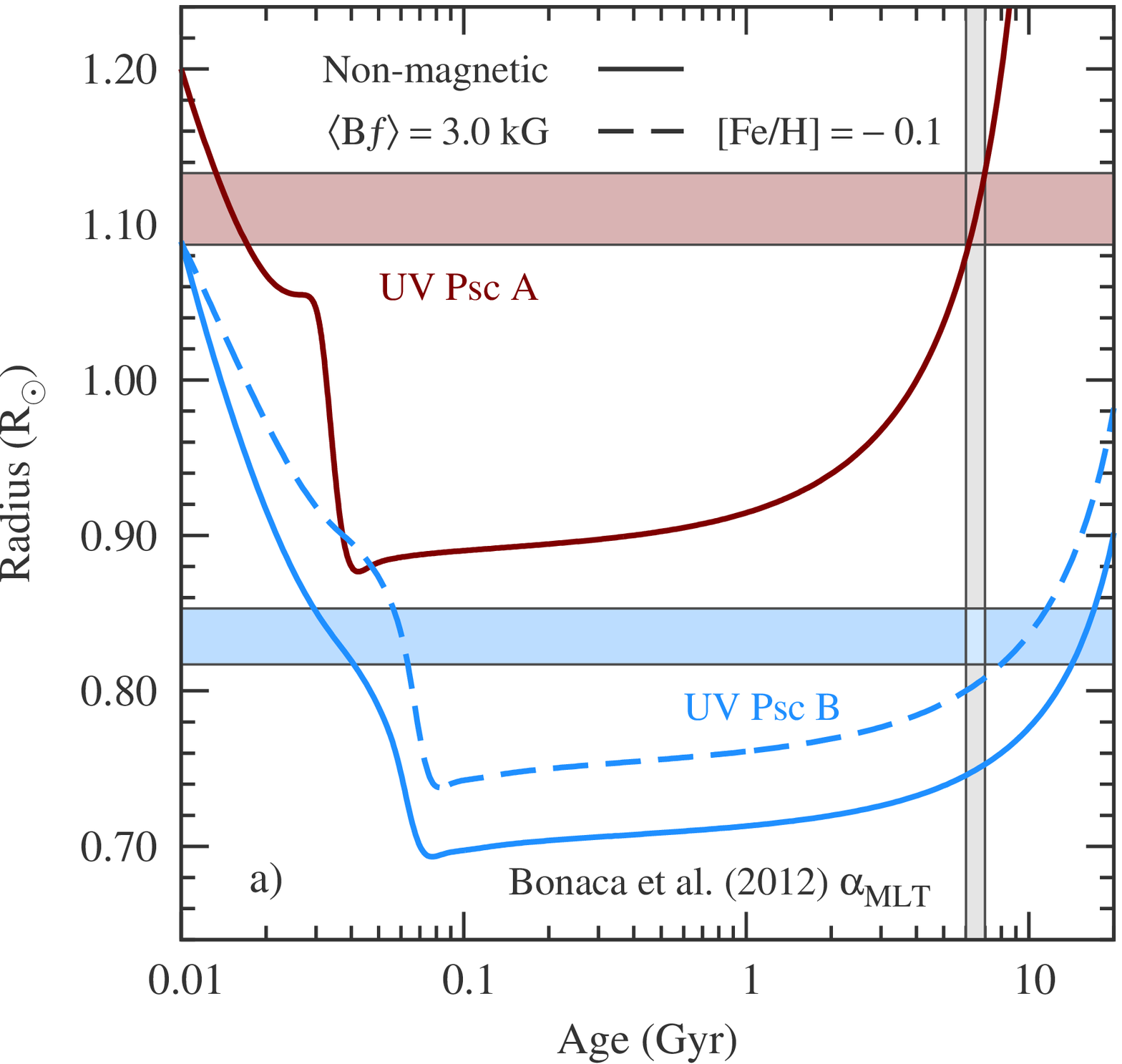}{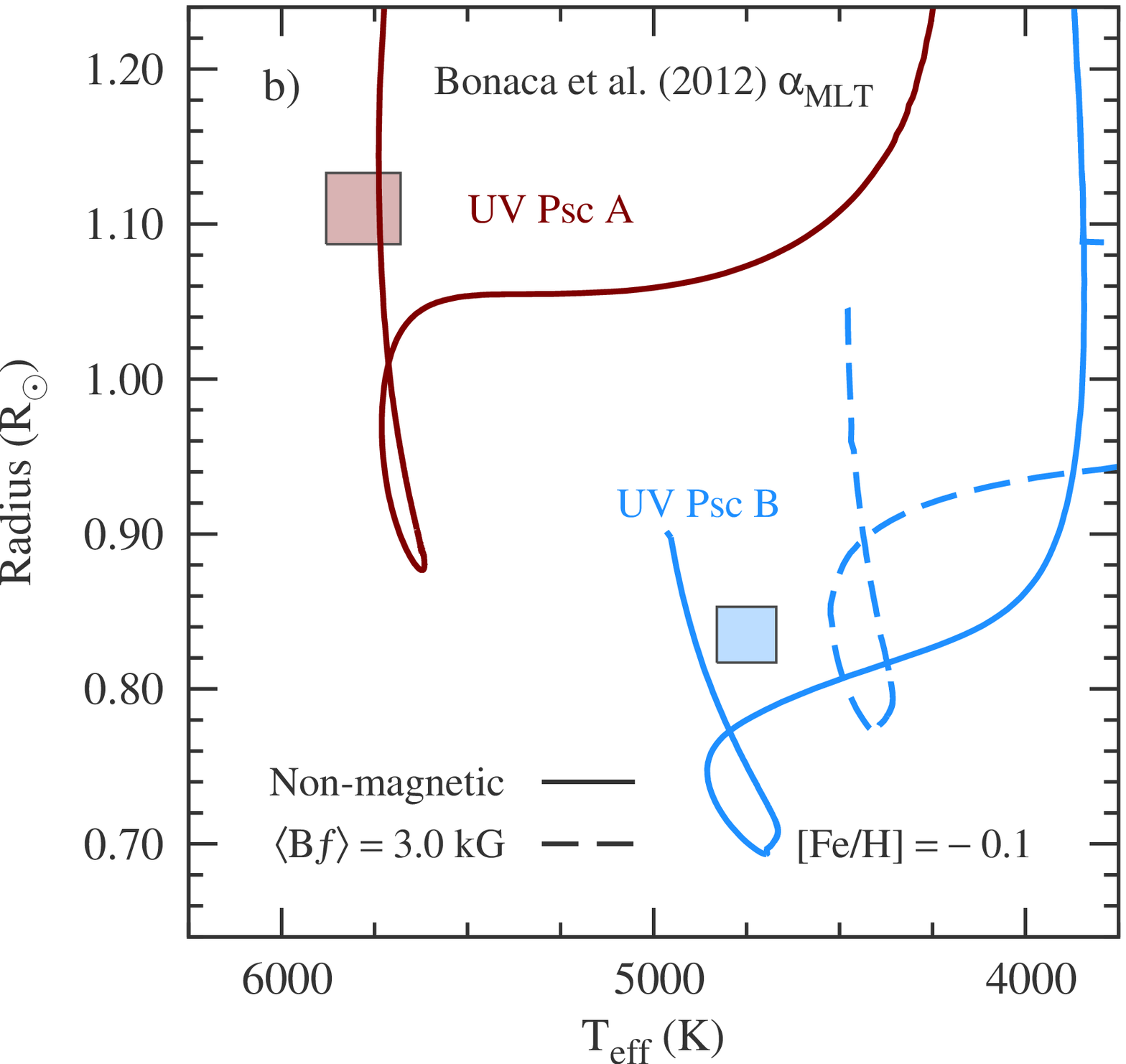}
    \caption
        {Similar to Figure \ref{fig:uvpsc_feh_b} except that all of the 
        mass tracks have a $\amlt$ reduced according to the \citet{Bonaca2012}
        empirical relation. The surface magnetic field strength used in
        modeling the secondary is $3.0$ kG. (a) Age-radius plane. (b)
        $\teff$-radius plane.
        \vspace{0.5\baselineskip}\\
        (A color version of this figure is available in the online journal.)
        }
    \label{fig:uvpsc_amlt}
\end{figure*}

Magnetic models of the secondary component were computed using a dipole 
profile, single-step perturbation at 10 Myr for several values of the 
surface magnetic field strength. 
A surface magnetic field strength of $4.0$ kG (corresponding to a tachocline 
field strength of $11$ kG) produced a model radius that was in agreement  
with the observed radius at 7.2\,Gyr. This 
is depicted in Figure~\ref{fig:uvpsc_feh_b}(a). The dashed line, representing
the magnetic model of the secondary, passes through the narrow region formed 
by the intersection of the radius (horizontal shaded area) and age 
(vertical shaded area) constraints.

We checked that the effective temperature predicted by the magnetic
model agreed with the temperature inferred from observations. Figure 
\ref{fig:uvpsc_feh_b}(b) shows the same 4.0 kG magnetic mass track required 
to fit the secondary in the age-radius plane over-suppresses the effective 
temperature. This causes the model to be too cool compared to the empirical 
value. Intuitively, one might suggest lowering the surface magnetic field 
strength so as to maintain agreement in the age-radius plane while allowing 
for a hotter effective temperature. However, all values of the surface
magnetic field strength that provide agreement in the age-radius plane 
produce models that are cooler than the empirical temperature.

How might we interpret the remaining temperature disagreement? One possible
solution is that the effective temperature measurement is incorrect. We 
feel this scenario is unlikely considering the temperatures are hot enough
where large uncertainties associated with complex molecular bands are not 
present. The uncertainties quoted in Table \ref{tab:deb_rad_core} seem 
large enough to encompass the actual value. Another possibility is that 
we have not treated convection properly. Convection within the component 
stars may not have the same inherent properties as convection within the 
Sun. This idea has continually motivated modelers to freely adjust the 
convective mixing length. However, while MLT is not 
entirely realistic and allows for such an arbitrary choice of the mixing 
length, arbitrary reduction without a definite physical motivation
(other than providing better empirical agreement) is not wholly 
satisfying. Glossing over the specific reasons for mixing length reduction 
does not fully illuminate the reasons for the noted discrepancies. 

Instead of applying an arbitrary adjustment to the convective mixing length, 
we modify the convective mixing length parameter according to the relation 
developed by \citet{Bonaca2012}. Using asteroseismic data, they provide 
a relation between the mixing length parameter and stellar physical properties 
(i.e., $\log g$, $\teff$, and [M/H]). Their formulation indicates that 
convection is less efficient (smaller mixing length) in low-mass, metal-poor
stars as compared to the solar case. Modifications to the convective mixing 
length are, therefore, no longer arbitrary and may not take on any value 
that happens to allow the models to fit a particular case. 

The \citet{Bonaca2012}
relation is based on models using an Eddington $T(\tau)$ relation to derive
the surface boundary conditions, meaning they require a solar calibrated 
mixing length of 1.69. Our use of {\sc phoenix} model atmosphere structures 
to derive the surface boundary conditions and treatment of atomic diffusion
of helium leads to our higher solar calibrated
mixing length of $\alpha_{\rm MLT,\, \odot} = 1.94$.  We therefore use the 
\citet{Bonaca2012} relation to derive the relative difference between the
empirically derived mixing length and their solar calibrated value, keeping
their fit coefficients fixed. New mixing lengths for
the stars in UV Psc are scaled from our solar mixing length using this 
relative difference. For a metallicity of $-0.1$\,dex, we find a mixing 
length of $\amlt = 1.71$ for the primary and $\amlt = 1.49$ for the secondary 
of UV Psc.

Resulting mass tracks are shown in Figure~\ref{fig:uvpsc_amlt}.
Directly altering convection in this manner does not provide an adequate 
solution. Reducing the mixing length inflates both of the stellar radii 
(Figure \ref{fig:uvpsc_amlt}(a)) and forces the temperature at the photosphere 
to decrease (see Figure \ref{fig:uvpsc_amlt}(b)). The mixing length primarily 
affects the outer layers of each star, where energy is transported by 
super-adiabatic convection. 
A lower mixing length implies that there is less energy flux across a given 
surface within the convection zone. Since  the star must remain in equilibrium, 
the outer layers puff up to increase the energy flux, thereby reducing the 
effective temperature.

\begin{figure*}[t]
    \plottwo{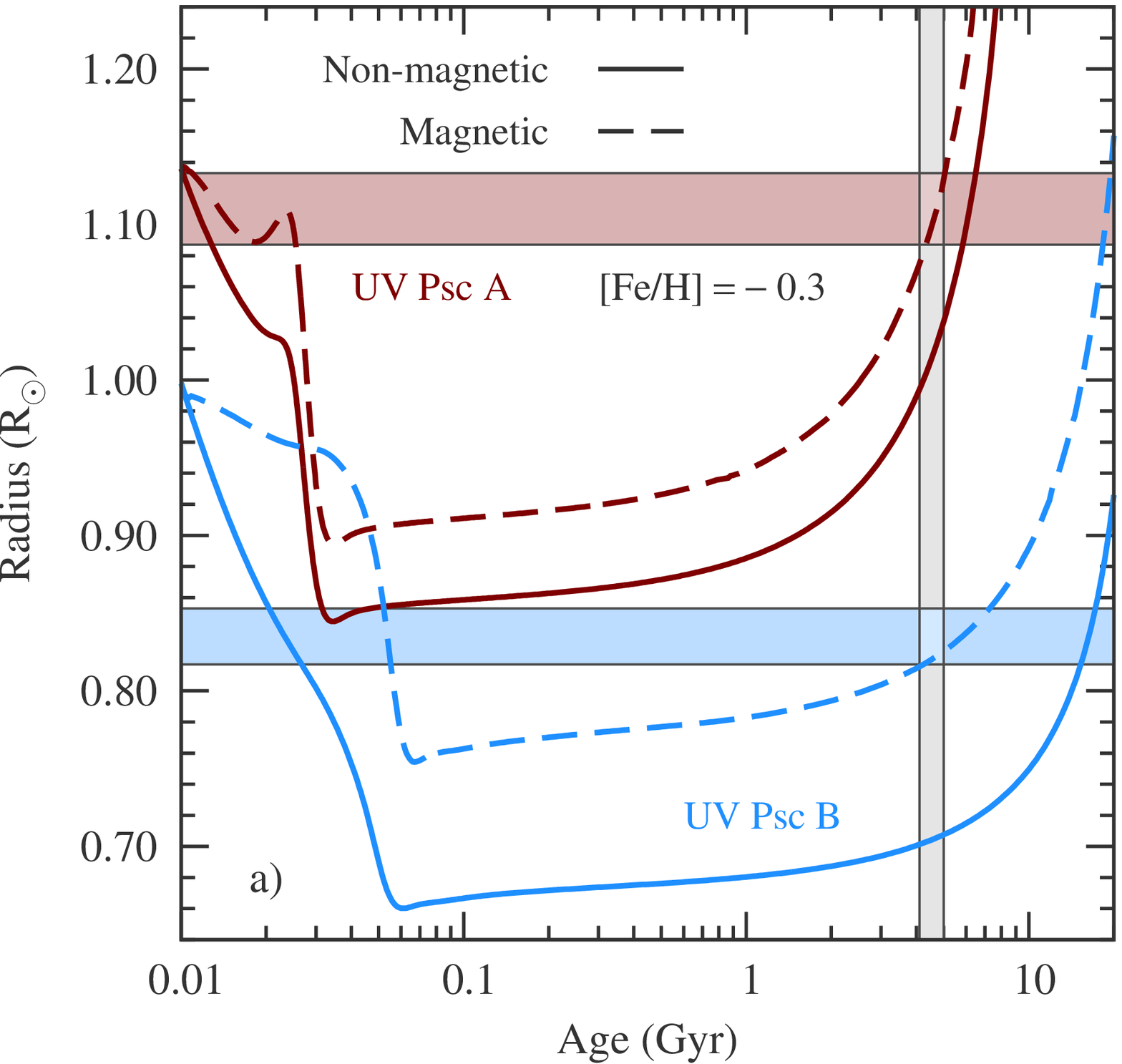}{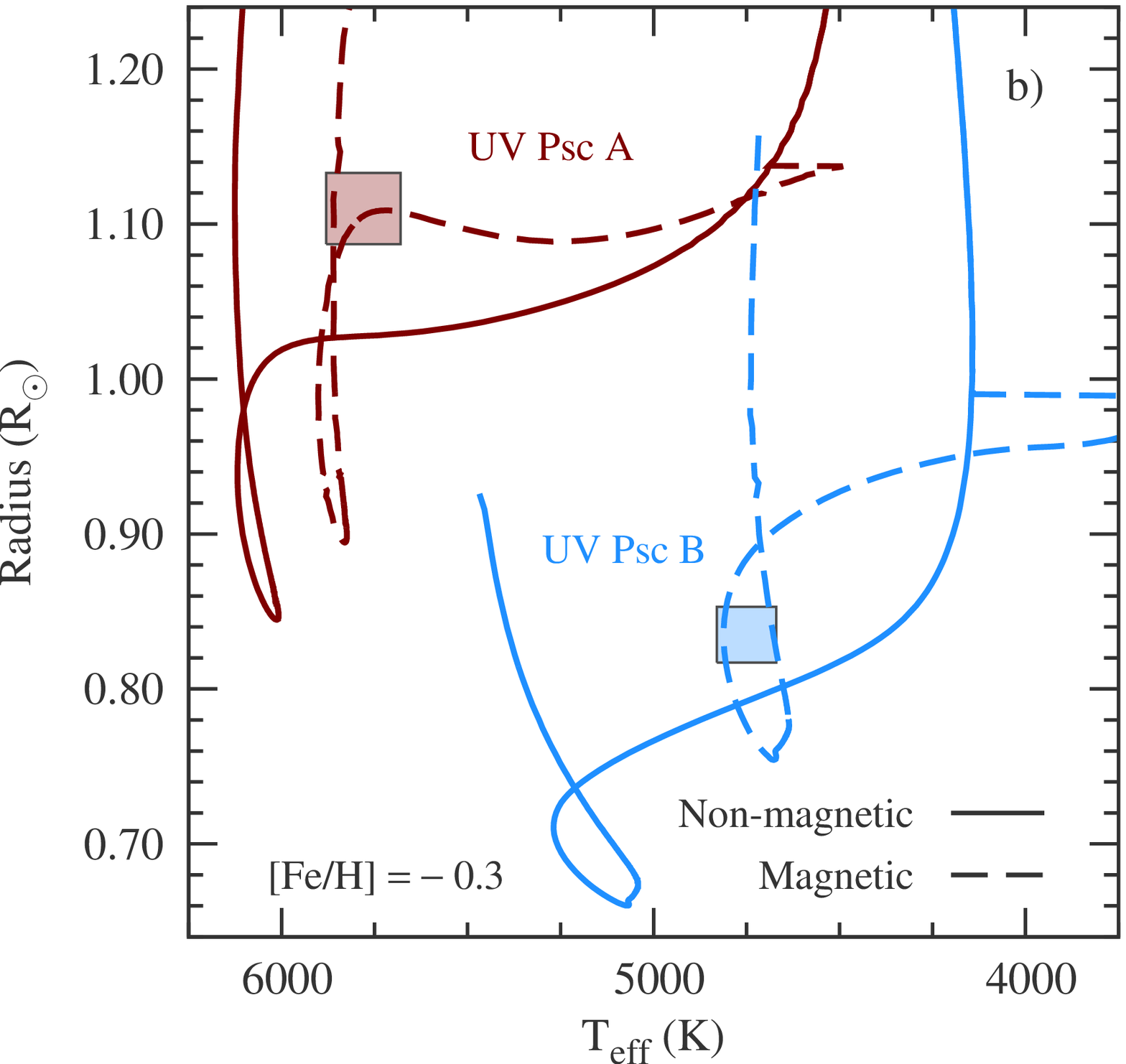}
    \caption{UV Psc system assuming a lower heavy 
        element abundance of [Fe/H] $= -0.3$. Standard DSEP mass tracks 
        are drawn as maroon and light blue solid lines for UV Psc A and
        B, respectively. Magnetic tracks are represented by dashed lines
        with the same color coding as the standard tracks. Surface magnetic
        field strengths are $2.0$ kG and $4.6$ kG for UV Psc A and B, 
        respectively. (a) Age-radius plane. (b) $\teff$-radius plane.
        \vspace{0.5\baselineskip}\\
        (A color version of this figure is available in the online journal.)
        }
    \label{fig:uvpsc_mag}
\end{figure*}

Inflating the primary component means the models of the secondary must now 
agree with the observed properties at an age younger than 7.2 Gyr. 
This is illustrated in Figure~\ref{fig:uvpsc_amlt}(a), where the 
vertical shaded area anchoring the system's age to UV Psc A is shifted 
left of where it was in Figure~\ref{fig:uvpsc_nmag}(a) by 0.5\,Gyr. 
A weaker magnetic field is now required to alleviate the radius 
disagreement with the secondary due to inflation caused by a reduced 
mixing length. Figure~\ref{fig:uvpsc_amlt}(a) shows a magnetic 
model with a surface magnetic field strength of 3.0\,kG. We do not find 
agreement between the model and empirical radius, but more importantly, 
Figure~\ref{fig:uvpsc_amlt}(b) demonstrates that the secondary's effective 
temperature is too cool. Increasing the surface magnetic field 
strength to produce agreement in the age-radius plane would only worsen 
the lack of agreement in the $\teff$-radius plane. Thus, reducing the 
mixing length is unable to provide relief to the magnetic over-suppression 
of the effective temperature in Figure
\ref{fig:uvpsc_feh_b}(b). We must seek another method to rectify the
effective temperature of the magnetic model.

Metallicity is an unconstrained input parameter for models of UV Psc. 
Recall, our selection of [Fe/H] $= -0.1$ was motivated by agreement 
of standard stellar evolution models with the primary. Updating our adopted
metallicity (and consequently, the helium abundance) has a non-negligible 
effect the structure and evolution of the UV Psc components. Stars with 
masses above $\sim0.45\msun$ are similarly affected by altering the 
chemical composition. For example, increasing the 
metallicity, and therefore the helium abundance, will increase the stellar 
radii and decrease the effective temperature. This is a result of changes 
to the {\it p-p} chain energy generation rate due to helium and 
the influence of both helium and heavy metals on bound-free radiative 
opacity. 

Adopting a lower metallicity of [Fe/H] $= -0.3$, while maintaining a solar
calibrated $\amlt=1.94$, for UV Psc increases the 
effective temperature of both standard model components and 
shrinks their radii at younger ages.\footnote{At older ages, evolutionary 
effects begin to play a role as stellar lifetimes are decreased at 
lower metallicity owing to higher temperatures within the stellar interior.}
Doing so also removes the effective temperature agreement between models 
and observations of UV Psc A. These effects are demonstrated for standard 
models in Figures~\ref{fig:uvpsc_nmag}(a) and (b), 
where we have plotted mass tracks with [Fe/H] $= -0.3$. Accurately 
reproducing the observed stellar properties now requires use of magnetic 
models for \emph{both} components.

Magnetic models with a dipole profile and single-step perturbation were 
constructed for both stars. We find that it is 
possible to wholly reconcile the models with the observations if the primary
has a $2.0$ kG surface magnetic field and the secondary has a $4.6$ kG surface
magnetic field. Model radii and temperatures match the empirical values 
within the age range specified by the primary, as shown in 
Figures~\ref{fig:uvpsc_mag}(a) and (b). 

The revised age of UV Psc found from Figure~\ref{fig:uvpsc_mag}(a)
(the vertical shaded region) is between 4.4 Gyr and 5.0 Gyr. Averaging 
the two implies an age of 4.7$\pm$0.3 Gyr. This age is nearly a factor 
of two younger than the 7 Gyr -- 8 Gyr age commonly prescribed to the system. 
While feedback from the models was necessary to adjust and improve upon 
the initial metallicity and to determine the required magnetic field strengths, 
we believe that this result is consistent with the available observational 
data. Our reliance on such a feedback cycle was inevitable given the lack 
of metallicity estimates. The metallicity range allowing for complete 
agreement is not limited to $-0.3$ dex. Further reducing the metallicity would
likely produce acceptable results, as the models
of UV Psc B just barely skirt the boundaries of the empirical values.
Our final recommendation is that UV Psc has a metallicity of  
[Fe/H] $= -0.3 \pm 0.1$ dex with surface magnetic fields of $2.0$ kG and $4.6$ kG
for the primary and secondary, respectively. Verification of these predictions
should be obtainable using spectroscopic methods.

\subsection{YY Geminorum}
\label{sec:yygem}

\begin{figure*}[t]
    \plottwo{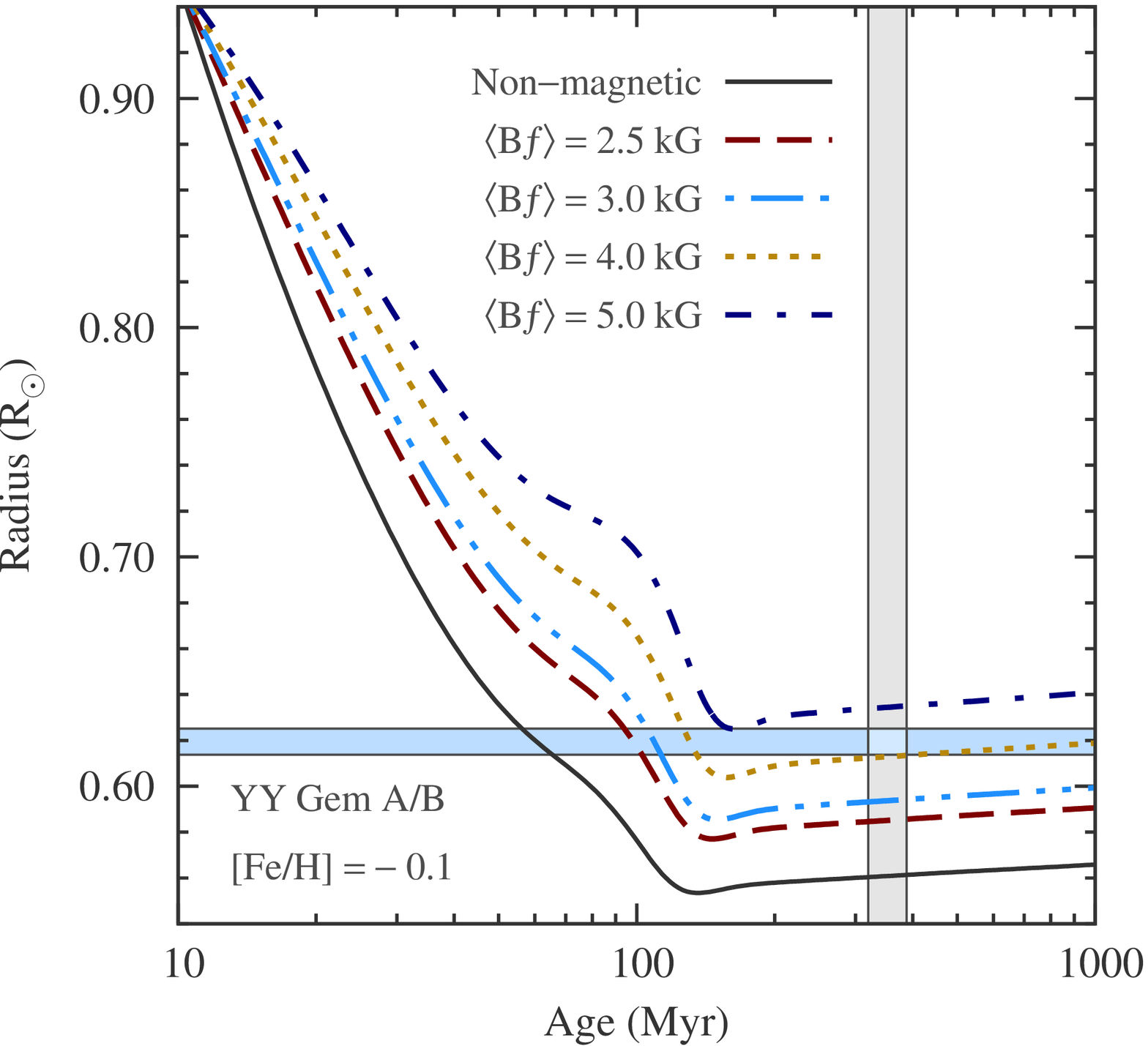}{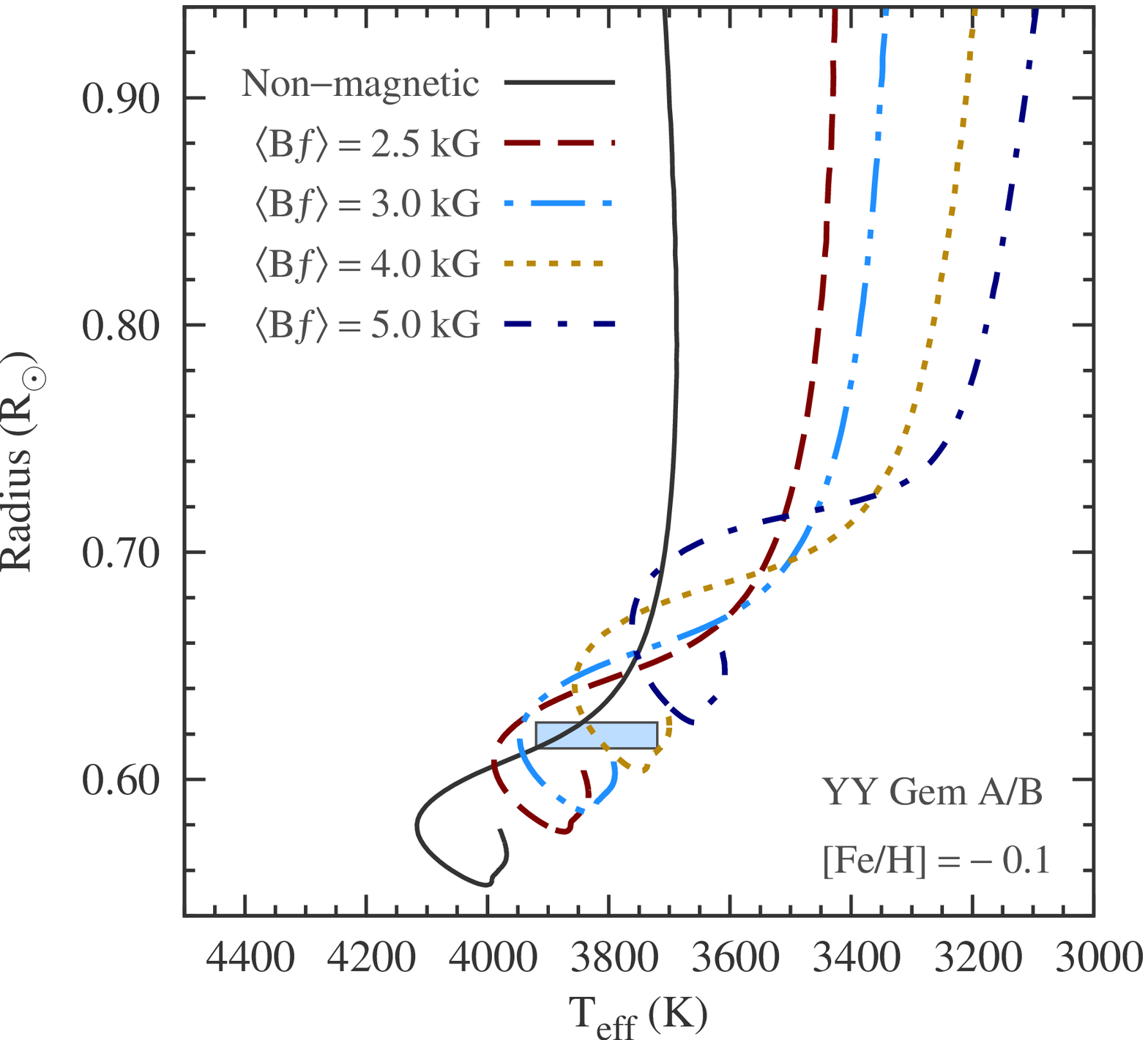}
    \caption{Standard (dark-gray, solid line) and magnetic stellar evolution
        mass tracks of YY Gem. Magnetic mass tracks were generated with 
        surface magnetic field strengths of $\langle {\rm B}f\rangle =$
        2.5 kG (maroon, dashed), 3.0 kG (light-blue, dash-double-dotted), 
        4.0 kG (mustard, dotted), and 5.0 kG (dark-blue, dash-dotted).
        All of the models were computed with a metallicity [Fe/H] $= -0.1$.
        Shown are the (a) age-radius plane, and (b) $\teff$-radius plane.
        The horizontal swaths represent the observational constraints in
        each plane while the vertical shaded region in panel (a) shows the
        estimated age constraints set by modeling of Castor A and B.
        \vspace{0.5\baselineskip}\\
        (A color version of this figure is available in the online journal.)
        }
    \label{fig:yygem_mr_dip}
\end{figure*}

YY Geminorum (also Castor C and GJ 278 CD; hereafter YY Gem) has 
been the subject of extensive investigation
after hints of its binary nature were spectroscopically uncovered \citep{Adams1920}. 
The first definitive reports of the orbit were published nearly simultaneously 
using spectroscopic \citep{Joy1926} and photographic methods \citep{vanGent1926},
which revealed the system to have an incredibly short period of 0.814 days.
Photographic study by \citet{vanGent1926} further revealed that the 
components eclipsed one another with the primary and secondary eclipse 
depths appearing nearly equal.
Rough estimates of the component masses and radii were carried out using 
the available data, but the data were not of sufficient quality to extract
reliable values \citep{Joy1926}. The system has since been confirmed to 
consist of two equal mass, early M-dwarfs. Masses and radii are now 
established with a precision of under 1\% \citep{Torres2002}. These measurements
are presented in Table~\ref{tab:deb_rad_core}.

The age and metallicity of YY Gem have been estimated using YY Gem's
common proper motion companions, Castor A and B. Considered gravitationally
bound, these three systems have been used to define the Castor moving group
\citep[CMG;][]{aoc89}. Spectroscopy of Castor Aa and Ba, both spectral-type A stars, 
yields a metallicity of [Fe/H] = +0.1 $\pm$ 0.2 \citep{Smith1974,Torres2002}.
Stellar evolution models of Castor Aa and Castor Ba provide an age estimate 
of 359 $\pm$ 34 Myr, which was obtained by combining estimates from multiple
stellar evolution codes \citep{Torres2002}, including the Dartmouth code
\citep{FC12}.

Over half a century after its binarity was uncovered, low-mass stellar 
evolution models suggested that the theoretically predicted radii may
not agree with observations \citep{Hoxie1970,Hoxie1973}. A subsequent 
generation of models appeared to find agreement with the observations 
\citep{CB1995}, but confirmation of the true discrepancies remained veiled 
by model and observational uncertainties. Modern observational determinations 
of the stellar properties \citep{Segransan2000,Torres2002} 
compared against sophisticated low-mass stellar evolution 
models \citep{BCAH98,Dotter2008} have now solidified that the 
components of YY Gem appear inflated by approximately 8\% \citep{Torres2002,FC12}.

Figure \ref{fig:yygem_mr_dip} shows a standard stellar evolution mass track 
for the components of YY Gem as a dark gray, solid line. We plot an $M = 0.599\msun$
mass with a metallicity of [Fe/H] $= -0.1$. 
That metallicity was found to provide good agreement to observational 
data by \citet{FC12}. The vertical shaded region highlights YY Gem's adopted 
age. Figure \ref{fig:yygem_mr_dip}(a) indicates that the standard model 
under predicts the radius measured by \citet{Torres2002} (illustrated by 
the horizontal shaded region) by about 8\%, within the required age range. 
Similarly, there is a 5\% discrepancy with the effective temperature shown in
Figure \ref{fig:yygem_mr_dip}(b). 

As a brief aside, it may be noted from Figure \ref{fig:yygem_mr_dip}(a) 
that our models are consistent with the properties of YY Gem around 60 Myr.
This age would imply that YY Gem has not yet settled onto the MS,
which occurs near an age of about 110 Myr. Previous studies have considered 
the possibility that YY Gem is still undergoing its pre-MS 
contraction \citep{CB1995,Torres2002} and provide mixed conclusions. However,
the more recent study by \citet{Torres2002} provides a detailed analysis 
of this consideration and concludes that it is erroneous to assume YY Gem 
is a pre-MS system. This is primarily due to YY Gem's association 
with the Castor quadruple. YY Gem is considered to be firmly on the main 
sequence, making the system discrepant with stellar models.

YY Gem exhibits numerous features indicative of intense magnetic activity. 
Light curve 
modulation has been continually observed \citep{Kron1952,Leung1978,Torres2002}, 
suggesting the presence of star spots. Debates linger about 
the precise latitudinal location and distribution \citep[e.g.,][]{Gudel2001}
of star spots, but spots contained below mid-latitude (between 45$^{\circ}$ 
and 50$^{\circ}$) 
appear to be favored \citep{Gudel2001,Hussain2012}. The components display
strong Balmer emission \citep{Young1989, Montes1995c} and X-ray emission 
\citep{Gudel2001,Stelzer2002,Lopezm2007,Hussain2012} during quiescence 
and have been observed to undergo frequent flaring events 
\citep{Doyle1990a,Doyle1990b,Hussain2012}. Furthermore, YY Gem has been 
identified as a source of radio emission, attributed to partially relativistic 
electron gyrosynchrotron radiation \citep{Gudel1993,McLean2012}.
Given this evidence, it is widely appreciated that the stars possess 
strong magnetic fields. Therefore, it is plausible to hypothesize that 
the interplay between convection and magnetic fields lies at the origin
of the model-observation disagreements.

We compute magnetic stellar evolution mass tracks with various
surface magnetic field strengths. The magnetic perturbation was included 
using a dipole magnetic field configuration and was added in a single
time step. These tracks are plotted in Figures \ref{fig:yygem_mr_dip}(a) 
and (b). We adopt a metallicity of $-0.1$ dex, 
consistent with our non-magnetic model. The level of radius inflation 
and temperature suppression increases as progressively stronger values 
of the surface magnetic field strength are applied. A 5.0 kG
surface magnetic field strength model over predicts the observed stellar
radii. Figure \ref{fig:yygem_mr_dip}(a) demonstrates that a surface
magnetic field strength of just over 4.0 kG is needed to reproduce the 
observed radii.

\begin{figure*}[t]
    \plottwo{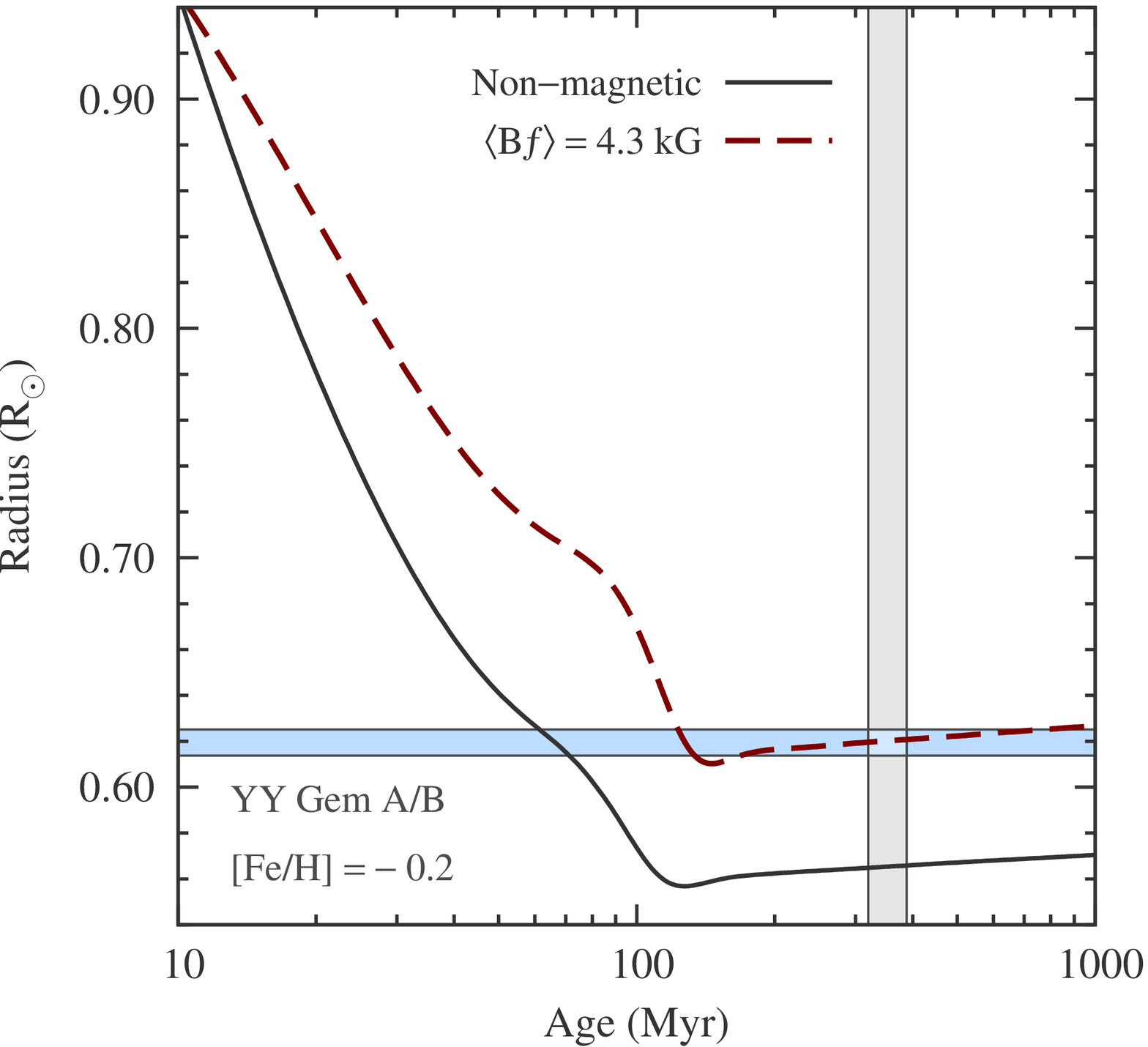}{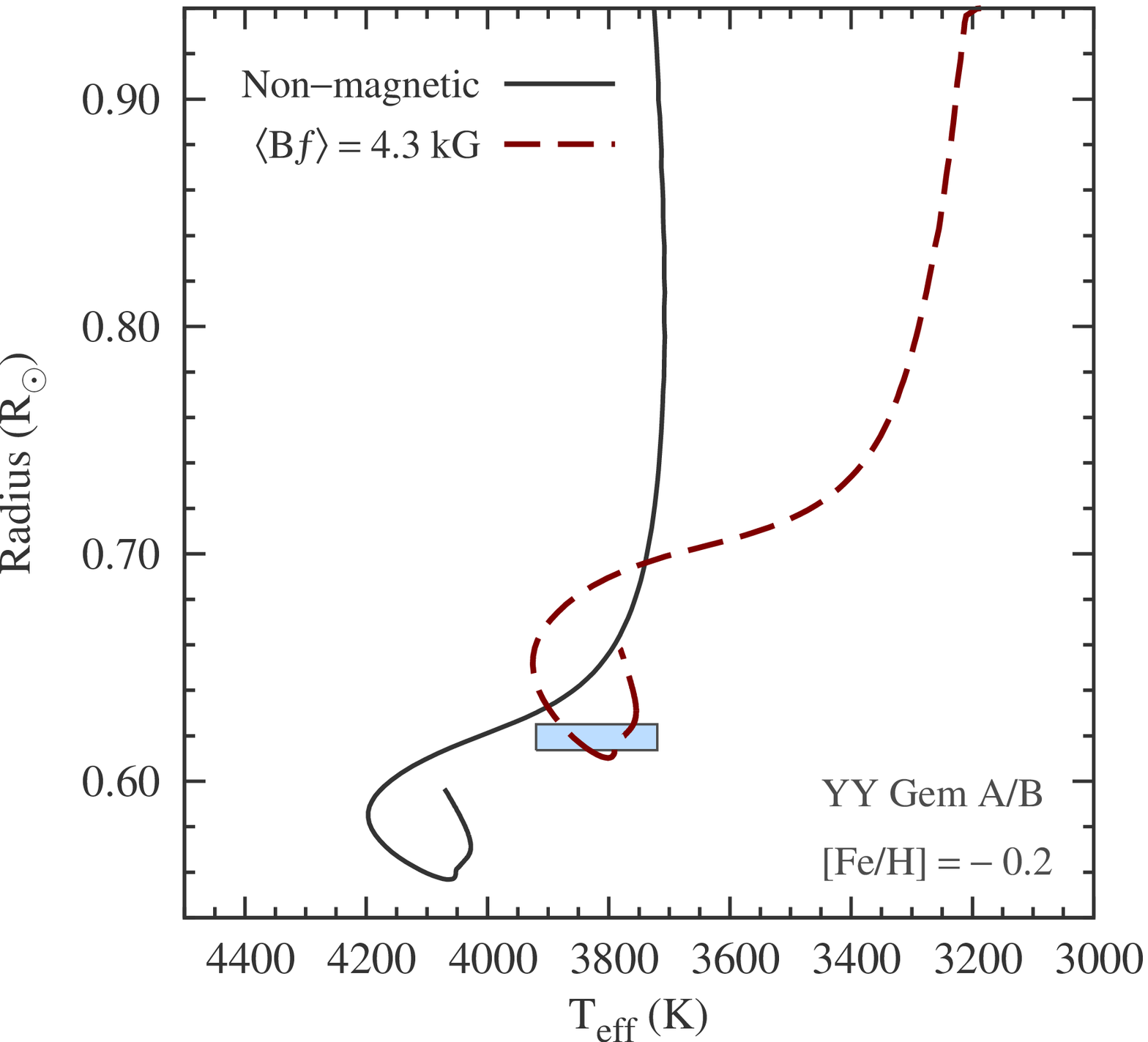}
    \caption{Standard (dark-gray, solid line) and magnetic stellar evolution
        mass tracks of YY Gem. The magnetic mass track was generated with 
        a surface magnetic field strength of $\langle {\rm B}f\rangle =$
        4.3 kG (maroon, dashed).
        Both of the models were computed with a metallicity [Fe/H] $= -0.2$.
        Shown are the (a) age-radius plane, and (b) $\teff$-radius plane.
        The horizontal swaths represent the observational constraints in
        each plane while the vertical shaded region in panel (a) shows the
        estimated age constraints set by modeling of Castor A and B.
        \vspace{0.5\baselineskip}\\
        (A color version of this figure is available in the online journal.)
        }
    \label{fig:yygem_mr_dip_m20}
\end{figure*}

Figure \ref{fig:yygem_mr_dip}(b) reveals that the models are barely able 
to match the observed effective temperature with a 4.0 kG magnetic field. 
Any stronger of a surface magnetic field over-suppresses the effective 
temperature, causing the model to be cooler than the observations. Recall, 
we encountered this same issue when attempting to model UV Psc in Section
\ref{sec:uvpsc}. A lower metallicity provides a solution for UV Psc, but 
doing so for YY Gem would jeopardize the metallicity prior established 
by the association with Castor AB \citep{Smith1974,Torres2002}. 

Before ruling out the option of a lower metallicity, we recompute 
the approximate metallicity of YY Gem using the \citet{Smith1974} values. 
First, we need to determine the metallicity of Vega, the reference for the
\citet{Smith1974} study. Vega has 21 listed metallicity measurements 
in SIMBAD, of which, the 8 most recent appear to be converge toward a 
common value. Using the entire list of 21 measurements, Vega has a metallicity 
of [Fe/H] $= -0.4\pm0.4$ dex. If, instead, we adopt only those measurements 
performed since 1980, we find [Fe/H] $= -0.6\pm0.1$ dex. The convergence
of values in recent years leads us to believe that this latter estimate
is more representative of Vega's metallicity.

Metallicities measured by \citet{Smith1974} for Castor A and Castor B 
were +0.98 dex and +0.45 dex, respectively. Averaging these two 
quantities as the metallicity for the Castor AB system, we have [Fe/H] 
$= +0.7\pm0.3$ dex. The difference in metallicity of Castor A and B might
be explained by diffusion processes \citep[e.g.,][]{Richer2000b} and is
not necessarily a concern. However, the fact that we are not observing the
initial abundances for Castor A and B is a concern when it comes to prescribing
a metallicity for YY Gem. Caution aside, a conservative estimate for the metallicity 
of YY Gem relative to the Sun is [Fe/H] $= +0.1\pm0.4$ dex. This new
estimate provides greater freedom in our model assessment of YY Gem. We 
note that this reassessment neglects internal errors associated
with the abundance determination performed by \citet{Smith1974}. Given the
large uncertainty quoted above, the real metallicity is presumed to lie 
within the statistical error. Confirmation of these abundances would be
extremely beneficial. New abundance determinations would not only enhance our 
understanding of YY Gem, but also provide evidence that the three 
binaries comprising the Castor system have a common origin.

Presented with greater freedom in modeling YY Gem, we compute additional
standard and magnetic mass tracks with [Fe/H] $= -0.2$ dex. The magnetic tracks
were computed in the same fashion as the previous set to provide a direct 
comparison on the effect of metallicity. Figures \ref{fig:yygem_mr_dip_m20}(a) 
and (b) illustrate the results of these models. 
Reducing the metallicity from [Fe/H] $= -0.1$ to [Fe/H] $= -0.2$ dex shrinks
the standard model radius by about 1\% at a given age along the MS.
As anticipated, a standard model with a revised metallicity also 
shows a 50\,K hotter effective temperature. 

A magnetic mass track with a surface magnetic field of 4.3 kG was found
to provide good agreement. At 360 Myr, it is apparent that the magnetic 
model of YY Gem satisfies the radius restrictions enforced by the observations.
The precise model radius inferred from the mass track is $0.620\rsun$, 
compared to the observed radius of $0.6194\rsun$, a difference of 0.1\%. 
Figure~\ref{fig:yygem_mr_dip_m20}(b) further demonstrates that when the 
model is consistent with the observed radius, the effective 
temperature of the mass track is in agreement with the observations. The 
model effective temperature at 360 Myr is 3773 K, well within the 1$\sigma$ 
observational uncertainty (also see Table~\ref{tab:deb_rad_core}).

There is one additional constraint that we have yet to mention. Lithium 
has been detected in the stars of YY Gem \citep{byn1997}. The authors 
find $\log N$(\el{7}{Li}) $ = 0.11$, where 
$\log N$(\el{7}{Li}) $= 12 + \log(X_{\rm Li}/A_{\rm Li}X_{\rm H})$. However,
standard stellar models predict that lithium is completely depleted 
from the surface after about 15 Myr---well before the stars reach the main 
sequence. Since magnetic fields can shrink the surface convection zone,
it is possible for the fields to extend the lithium depletion timescale 
\citep{MM10}. This is precisely what our magnetic models predict. With
a metallicity of [Fe/H] $= -0.2$ and a 4.3 kG surface magnetic
field our models predict $\log N$(\el{7}{Li}) $ \sim 0.9$ at 360 Myr. With 
[Fe/H] $= -0.1$ and a $4.0$ kG we find $\log N$(\el{7}{Li}) $= 0.1$ at 
360 Myr. The latter value is consistent with the lithium abundance 
determination of \citet{byn1997}, but is inconsistent with the metallicity 
motivated by agreement with the fundamental stellar properties.

In summary, we find good agreement with magnetic models that have a surface
magnetic field strength between $4.0$ and $4.5$ kG. A sub-solar metallicity 
of [Fe/H] $= -0.2$ provides the most robust fit with fundamental properties, 
but a metallicity as high as [Fe/H] $= -0.1$ may be allowed. The 
latter metallicity provides a theoretical lithium abundance estimate 
consistent with observations. A lower metallicity model predicts too much 
lithium at 360 Myr. It should be possible to confirm each of these
conclusions observationally.

\subsection{CU Cancri}
\label{sec:cucnc1}

The variable M-dwarf CU Cancri \citep[GJ 2069A, hereafter CU Cnc;][]{Haro1975} 
was discovered to be a double-lined spectroscopic binary 
\citep{Delfosse1998}. Follow up observations provided evidence that 
CU Cnc underwent periodic eclipses, making it the third known M-dwarf DEB
at the time \citep{Delfosse1999}. Shortly thereafter, \citet{Ribas2003}
obtained high-precision light curves in multiple photometric passbands.
Combining his light curve data and the radial velocity data from \citet{
Delfosse1998}, \citet{Ribas2003} published a detailed reanalysis of CU 
Cnc with precise masses and radii for the two component stars. These values 
are presented in Table \ref{tab:deb_rad_core}.

Initial comparisons with \citet{BCAH98} solar metallicity models indicated 
that the components of CU Cnc were 1 mag under luminous in the $V$ 
band. Additionally, the prescribed spectral type was two subclasses later than 
expected for two $0.4\msun$ stars \citep[M4 instead of M2;][]{Delfosse1999}. 
These oddities provided evidence that CU Cnc may have a super-solar metallicity.
An increased metallicity would increase TiO opacity at optical wavelengths
producing stronger TiO absorption features used for spectral classification. 
Absolute $V$ band magnitudes would also be lowered since TiO bands primarily 
affect the opacity at optical wavelengths, shifting flux from the optical 
to the near-infrared. Using \citet{BCAH98} models with metallicity $0.0$ 
and $-0.5$, \citet{Delfosse1999} performed a linear extrapolation to 
estimate a metallicity of [Fe/H] $\sim +0.5$.

A super-solar metallicity, as quoted by \citet{Delfosse1999}, is supported 
by the space velocity of CU Cnc. It has galactic velocities $U \approx -9.99 
\textrm{ km s}^{-1}$, $V \approx -4.66 \textrm{ km s}^{-1}$, and $W \approx 
-10.1 \textrm{ km s}^{-1}$ and is posited to be a member of the thin-disk
population. This population is characterized by younger, more metal-rich 
stars. However, space velocities were used by \citet{Ribas2003} to refute 
the \citet{Delfosse1999} metallicity estimate. Instead of indicating that 
CU Cnc has a super-solar metallicity, \citeauthor{Ribas2003} conjectured 
that the space velocities of CU Cnc implied it was a member of the 
CMG. The CMG is defined by $U = -10.6\pm3.7 \textrm{ km s}^{-1}$, 
$V = -6.8\pm2.3 \textrm{ km s}^{-1}$, and $W = -9.4\pm2.1 \textrm{ km s}^{-1}$.
Therefore, \citeauthor{Ribas2003} prescribed the metallicity of the Castor
system to CU Cnc (see Section \ref{sec:yygem}), suggesting that CU Cnc may
have a near-solar or slightly sub-solar metallicity. 

With a metallicity and age estimate defined by the CMG, 
\citet{Ribas2003} performed a detailed comparison between stellar models 
and the observed properties of CU Cnc. Models of the CU Cnc stars
were found to predict radii 10\% -- 14\% smaller than observed. 
Furthermore, effective temperatures were 10\% -- 15\% 
hotter than the effective temperatures estimated by \citet{Ribas2003}. 
CU Cnc was found to be under luminous in the $V$ and $K$ band by
1.4 mag and 0.4 mag, respectively. \citeauthor{Ribas2003}
proceeded to lay out detailed arguments that neither stellar activity
nor metallicity provides a satisfactory explanation for the observed radius,
$\teff$, and luminosity discrepancies. Instead, he proposes that CU Cnc
may possess a circumstellar disk. The disk would then disproportionately 
affect the observed $V$ band flux compared to the $K$ band. This would also
force the effective temperatures to be reconsidered, leading to a change
in the observed luminosities. 
 
\citet{Ribas2003} relies heavily on the estimated effective
temperature of the individual components. Determining M-dwarf effective 
temperatures is fraught with difficulty. There is a strong
degeneracy between metallicity and effective temperature for M-dwarfs
when considering photometric color indices. We will therefore return 
to a detailed discussion of the luminosity discrepancies later and focus
on the radius deviations first. Radius estimates will be less affected by 
the presence of a circumstellar disk since radius determinations rely on 
differential photometry.

In \citet{FC12}, our models preferred a super-solar metallicity 
when attempting to fit CU Cnc. The maximum metallicity permitted in that 
analysis was [Fe/H] $= +0.2$ dex. Since CU Cnc may have a metallicity 
greater than the limit in \citet{FC12}, we begin with a standard model 
analysis of CU Cnc assuming a super-solar metallicity with [Fe/H] $\ge 
+0.2$ dex. Allowing for CU Cnc to have a super-solar metallicity, or in 
particular a metallicity different from YY Gem, contradicts its proposed 
membership with the CMG. However, even though CU Cnc has a similar velocity
to Castor (within 3~km~s$^{-1}$), other proposed members of the CMG have
been shown to differ significantly from Castor (and each other) in their 
velocities \citep{Mamajek2013}. \citet{Mamajek2013} present detailed arguments
that show the motions of CMG members are dominated by the Galactic potential,
meaning members very likely do not have a common birth site. While CU Cnc 
may have common properties with Castor, it is far from certain whether the two
share a common origin. Therefore, we reject the CMG association, thus 
allowing for age and metallicity to be free parameters in our modeling.

\begin{figure}[t]
    \plotone{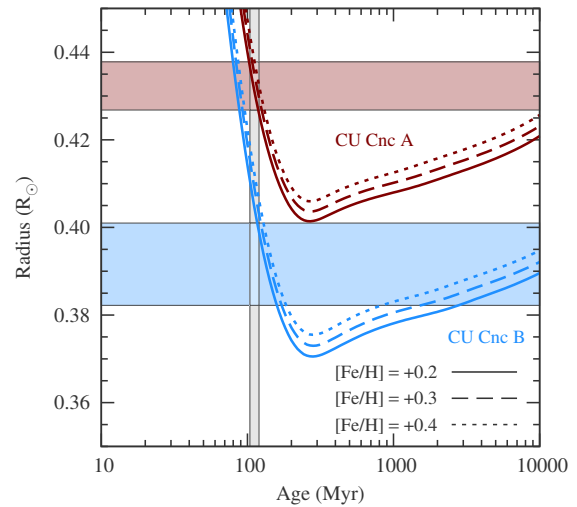}
    \caption{Standard Dartmouth mass tracks for CU Cnc A (maroon) and 
        CU Cnc B (light-blue) at three different metallicities: $+0.2$
        (solid), $+0.3$ (dashed), $+0.4$ (dotted). Horizontal bands identify
        the observed radius with $1\sigma$ uncertainty while the vertical
        band identifies the region in age-radius space where the models 
        match the primary star's observed radius.
        \vspace{0.5\baselineskip}\\
        (A color version of this figure is available in the online journal.)}
    \label{fig:cu_cnc_nmag}
\end{figure}

Standard stellar evolution models of both components are presented in Figure 
\ref{fig:cu_cnc_nmag}. Results are nearly independent of the adopted 
metallicity. All mass tracks show that the models do not match the observed 
stellar radii at the same age along the MS. Models of the primary 
appear to deviate from the observations more than models of the secondary. 
This may just be a consequence of the larger radius uncertainty quoted for 
the secondary star, creating an illusion of better agreement.  Quoting 
precise values for the level of disagreement is difficult as it depends 
strongly on the adopted age. Assuming an age of 360\,Myr, our models under 
predict the radius of the primary and secondary by 7\% and 5\%, respectively.

\begin{figure}[t]
    \plotone{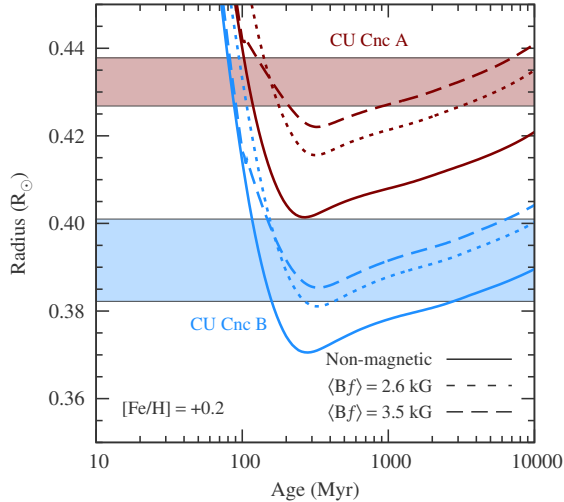}
    \caption{Magnetic stellar evolution mass tracks of CU Cnc A (maroon)
        and CU Cnc B (light-blue) with surface field strengths of 2.6 kG 
        (dotted) and 3.5 kG (dashed). A non-magnetic mass track for each
        star is shown as a solid line. All models have [Fe/H] $ = +0.2$
        following the discussion in the text. The horizontal swaths signify
        the observed radius with associated $1\sigma$ uncertainty. 
        \vspace{0.1\baselineskip}\\
        (A color version of this figure is available in the online journal.)}
    \label{fig:cu_cnc_mag}
\end{figure}

Agreement between the models and observations for both components is seen
near 120\,Myr (vertical shaded region in Figure \ref{fig:cu_cnc_nmag}). At 
this age, the stars are undergoing gravitational contraction along the 
pre-main-sequence (pre-MS). We cannot rule out the possibility that the stars 
of CU Cnc are still in the pre-MS phase. \citet{Ribas2003} 
tentatively detects lithium in the spectrum of CU Cnc, which strongly suggests
it is a pre-MS system. However, models predict complete lithium depletion
around 20\,Myr, 100\,Myr prior to where the models show agreement. This is
almost entirely independent of metallicity.  Only by drastically lowering 
the metallicity to $-1.0$\,dex are we able to preserve some lithium at the
surface of CU Cnc A as it reaches the MS. We note 
also that agreement between the models and observations occurs right at 
the edge of the gray vertical area in Figure \ref{fig:cu_cnc_nmag}, suggesting 
that the agreement may be spurious. 

For the purposes of this study, we assume that the stars have reached the
MS and that magnetic fields may underlie the observed radius
discrepancies. There is evidence that the stars are magnetically active. 
{\it ROSAT} observations show strong X-ray emission\footnote{{\it ROSAT}
observations actually contain emission from both CU Cnc and its proper motion
companion, the spectroscopic binary
CV Cnc.} \citep{Lopezm2007,FC12} indicative of the stars having magnetically 
heated coronae. CU Cnc is also classified as an optical flare star
that undergoes frequent flaring events \citep{Haro1975, Qian2012}.
Furthermore, the stars show strong chromospheric Balmer and Ca {\sc ii} 
K emission during quiescence \citep{Reid1995,Walkowicz2009}. These tracers 
point toward the presence of at least a moderate level of magnetic activity 
on the stellar surfaces. 

Magnetic models were computed using a dipole magnetic field profile and 
two surface magnetic field strengths were chosen, 2.6\,kG and 3.5\,kG.
Mass tracks including a magnetic field are shown in Figure \ref{fig:cu_cnc_mag}.
We fixed the metallicity to [Fe/H] $ = +0.2$ since it makes only a 
marginal difference in the overall radius evolution of standard model mass 
tracks. Note that the magnetic perturbation time is different between the 
2.6\,kG and 3.5\,kG tracks. The perturbation age was pushed to 100\,Myr 
when using a 3.5\,kG model to ensure model convergence immediately following 
the perturbation. We performed numerical tests to confirm that altering 
the perturbation age does not influence results along the MS.

Figure \ref{fig:cu_cnc_mag} shows that 
our model of the secondary star with a surface magnetic field strength of
3.5\,kG matches the observed radius between 300\,Myr and 6\,Gyr
(ignoring the pre-MS). A lower, 2.5\,kG, surface magnetic 
field strength produces similar results, but shows slight disagreement 
with the observations near the zero-age main sequence (ZAMS) at 300\,Myr.
However, the 2.5\,kG model extends the maximum age from 6\,Gyr to 10\,Gyr.
Unlike models of the secondary, neither of the magnetic models of the primary 
produce agreement near the ZAMS. Instead, agreement is obtained between
900\,Myr and 6\,Gyr. To create agreement between the model and observed radius
near the ZAMS, our models would require a stronger surface magnetic field 
strength.

\begin{figure}[t]
    \plotone{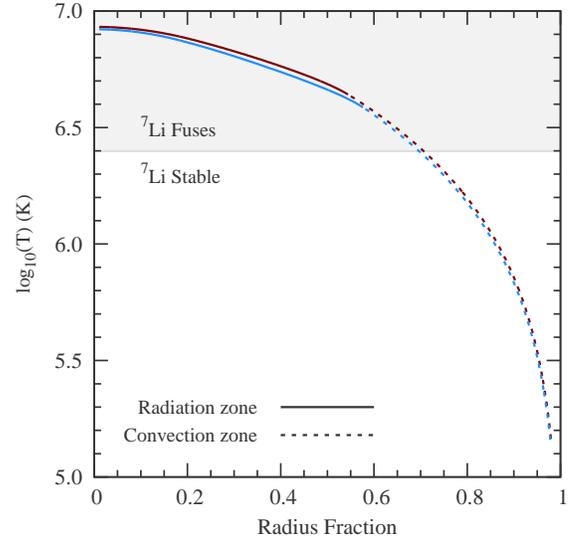}
    \caption{Temperature profile within a standard (maroon) and magnetic 
        (light-blue) model of CU Cnc A showing that the
        base of the convection zone exists at a higher temperature than
        the $^7$Li fusion temperature (gray shaded area above $\log(T) = 6.4$). 
        The influence of the magnetic field on the location of the convection
        zone base can be seen.
        Note that the temperature profile from the stellar envelope 
        calculation is not included.
        \vspace{0.5\baselineskip}\\
        (A color version of this figure is available in the online journal.)}
    \label{fig:li7_burn}
\end{figure}

The need for a stronger magnetic field in the primary depends on the real 
age of the system. \citet{Ribas2003} invoked the possible CMG
membership to estimate an age. The CMG is thought 
to be approximately 350 -- 400 Myr (see Section \ref{sec:yygem}). According 
to Figure \ref{fig:cu_cnc_mag}, this would place CU Cnc near the ZAMS.
It also means that a stronger magnetic field would be needed in modeling 
the primary star. In fact, a surface magnetic field strength of 4.0 kG is 
required to produce agreement with the primary if CU Cnc is coeval with 
the CMG. However, there is no compelling argument that leads us 
to believe that CU Cnc has properties in common with Castor. Kinematic 
similarities among field stars is not sufficient for assigning a 
reliable age or metallicity \citep[e.g.,][]{Mamajek2013}. 

If CU Cnc is a young system near the ZAMS, a magnetic field
may hinder lithium depletion from the stellar surface. We saw that 
this occurred with YY Gem in Section \ref{sec:yygem}. However, depletion 
of lithium from the surface of the stars in CU Cnc is unaffected by a 
strong magnetic field. Unlike YY Gem, where lithium was preserved 
to a significantly older age, the stars in CU Cnc destroy lithium along 
the pre-MS, nearly independent of metallicity and magnetic field strength. 
This can be readily 
explained by the depth of the convective envelope in the stars of CU Cnc. 
At $\sim0.4\msun$, the stars are expected to have deep convection zones 
that extend from the stellar surface down to about 55\% of the stellar 
radius. Lithium is destroyed ($T = 2.5\times10^6$ K) at a depth
located considerably closer to the stellar surface than the convection
zone boundary. This is illustrated in Figure \ref{fig:li7_burn}. 
Lithium will be mixed down to the base of the convection zone where it 
will rapidly burn, leading to a complete absence of lithium by 20\,Myr. 
Introducing a strong magnetic field (3.5 kG) reduces the size 
of the convection zone in CU Cnc A by 7\% and by 9\% in CU Cnc B. In contrast,
to preserve lithium the size of the convection zone would have to be reduced
by nearly 30\%. This corresponds to the base of the convection zone moving
from 55\% to 70\% of the total stellar radius. If lithium exists at 
the surface of CU Cnc, then there is another process keeping lithium from 
being destroyed. 

We conclude this section on CU Cnc by returning to the photometric issues
raised by both \citet{Delfosse1999} and \citet{Ribas2003}. Since the 
publication of \citet{Ribas2003}, \emph{Hipparcos} parallaxes have
been revised and updated to provide more accurate solutions \citep{hip07}.
The parallax for CU Cnc underwent a revision from $\pi = 78.05 \pm 5.69$
mas to $\pi = 90.37 \pm 8.22$ mas, changing the distance estimate from 
$12.81\pm0.92$ pc to $11.07\pm1.01$ pc. Absolute magnitudes must be
adjusted for this revised distance. Using $V$- and $K$-band magnitudes 
listed in \citet{Weis1991}\footnote{The integrated $V$-band magnitude 
listed on SIMBAD is 0.2 mag fainter than is quoted by \citet{Weis1991}. 
This is probably because the photometry listed on SIMBAD was taken during 
an eclipse, where the $V$-band flux drops by $\sim0.2$ mag \citep{Ribas2003}.} 
and the Two Micron All Sky Survey (2MASS) archive \citep{2mass}, we find 
$M_{V,\, A} = 12.27$ mag, $M_{V,\, B} = 12.63$ mag, and an integrated 
$M_K = 6.382$ mag. From SIMBAD we obtain integrated colors: $(J-K) = 0.906$ 
and $(H-K) = 0.291$, drawn from the 2MASS survey.

Using theoretical color-$\teff$ transformations \citep{Dotter2007,Dotter2008} 
we convert model surface properties to photometric magnitudes and colors.
We were unable to reproduce the set of integrated colors and magnitudes
or the individual $V$-band magnitudes using super-solar metallicity models
alone. However, combining a super-solar metallicity with a magnetic field,
we were able to produce models showing the appropriate trends: total $V$-
and $K$- band magnitudes were reduced due to the decrease in luminosity 
associated with a magnetic field. There was a steeper decrease in the $V$-band
due to increased metallicity and decreased $\teff$ due
to both metallicity and a magnetic field. The final photometric properties
of our models did not exactly match the properties of CU Cnc. We
note there is considerable uncertainty in the color-$\teff$ transformation
using the {\sc phoenix ames-cond} theoretical models \citep{BCAH98,Delfosse1998},
particularly in the $V$-band. A larger exploration of the model parameter
space and upgrading to the latest {\sc phoenix bt-settl} models would 
help to determine if metallicity and magnetic fields are able to resolve
the CU Cnc photometric anomalies. Whether a dusty disk exists should be testable 
using photometric data from {\it WISE} \citep{WISE}, which may reveal excess infrared 
emission.

\section{Magnetic Field Strengths}
\label{sec:mag_strengths}

Section \ref{sec:ind_deb} demonstrates that introducing a magnetic perturbation 
within stellar models can reconcile the model predictions with observations
of low-mass stars in DEBs. But, the real predictive power of the models 
relies on their ability to do so with realistic magnetic field strengths.
Better yet, with magnetic field strengths that translate into physical
observables. In \citet{FC12b}, we showed that it was possible to test the
validity of magnetic models using the stellar X-ray luminosity. Magnetic
models of EF Aquarii appeared to be consistent with this analysis. We 
perform a similar analysis for the stars studied in this paper.

\begin{deluxetable*}{l c c c c c}[]
    \tablewidth{1.5\columnwidth}
    \tablecaption{X-ray properties for the Three DEB systems.}
    \tablehead{
        \colhead{DEB} & \colhead{$X_{\rm cr}$} & \colhead{HR} & \colhead{$\pi$} &
        \colhead{$N_{\rm stars}$}  &  \colhead{$L_{x}$} \\
        \colhead{System} & \colhead{$(\textrm{cts s}^{-1})$} & & \colhead{(mas)} &
        \colhead{}  & \colhead{$(\textrm{erg s}^{-1})$}
    }
    \startdata
    UV Psc &  $0.92\pm0.07$  &  $-0.10\pm0.07$  &  $14.64\pm1.03$ & $2$ &  $(2.0\pm0.3)\times10^{30}$ \\
    YY Gem &  $3.70\pm0.09$  &  $-0.15\pm0.02$  &  $d=13\pm2$ pc  & $6$ &  $(9.4\pm0.2)\times10^{28}$ \\
    CU Cnc &  $0.73\pm0.05$  &  $-0.14\pm0.06$  &  $90.37\pm8.22$ & $4$ &  $(2.0\pm0.2)\times10^{28}$
    \enddata
    \tablecomments{$N_{\rm stars}$ is the total number of stars thought 
    to be contributing to the total X-ray counts detected by ROSAT. The 
    value $L_{x}$ is quoted as the X-ray luminosity per star in the system.}
    \label{tab:rad_rosat}
\end{deluxetable*}

\subsection{Estimating Surface Magnetic Field Strengths}
\label{sec:surf_field_est}

Estimates of surface magnetic field strengths can be obtained using an
empirical 
scaling relation derived between the total X-ray luminosity ($L_x$) of a 
star and its surface magnetic flux ($\Phi$) \citep{Fisher1998,Pevtsov2003}.
The relation states that \citep{Pevtsov2003}
\begin{equation}
    L_x \propto \Phi^{p},
\end{equation}
where $p = 1.15$, and appears to extend over 12 orders of magnitude in 
$\Phi$ and $L_x$. This includes data from individual solar quiet regions, 
solar active regions, and hemispherical averages from single field stars,
among others. However, the adopted power-law index does show variation 
between the individual data sets. For example, \citet{Pevtsov2003} found 
a power-law index of $p=0.98$ when they only considered dwarf star data 
from \citet{Saar1996}. This is quite different from the power-law index 
of $p=1.15$ derived when analyzing the ensemble of observations.

\begin{figure}
    \plotone{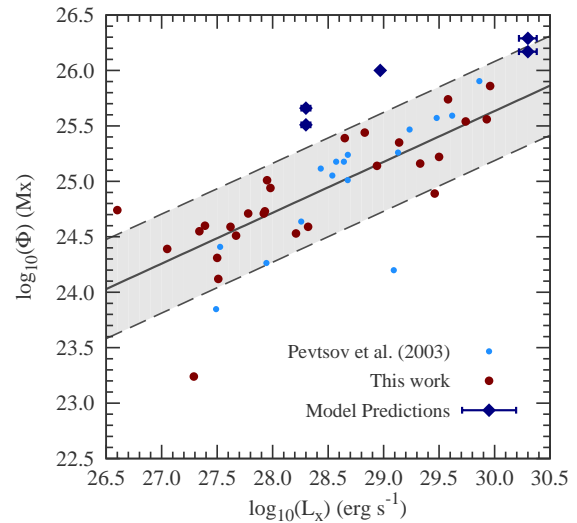}
    \caption{Unsigned stellar surface magnetic flux ($\Phi$) as a function 
         of the total stellar X-ray luminosity for a collection of G-, K-, 
         and M-dwarfs. Data shown are from \citet{Pevtsov2003} (small, gray
         points), this work (large, maroon points), and theoretical predictions
         from this work (blue, diamonds). 
         \vspace{0.5\baselineskip}\\
        (A color version of this figure is available in the online journal.)
        \label{fig:lx_mx_rad}}   
\end{figure}

The question of which power law index to adopt is important and can strongly 
influence the magnetic flux derived from X-ray luminosities. Seeing as 
we are focused on deriving the approximate magnetic flux for dwarf stars,
it seems a natural choice to use the relation specifically derived from 
dwarf star data. The relation derived by \citet{Pevtsov2003} was based 
on magnetic field measurements presented by \citet{Saar1996}. Since then,
many more stars have had their surface magnetic field strengths measured
\citep[see][for reviews]{Donati2009,Reiners2012a}. 

\begin{deluxetable}{l c c c}[t]
    \tablewidth{\columnwidth}
    \tablecaption{Surface Magnetic Field Properties for 
     UV~Psc, YY~Gem, and CU~Cnc}
    \tablehead{
    \colhead{DEB} & \colhead{$\log\Phi$} & \colhead{$\langle Bf\rangle$} & 
    \colhead{$\langle Bf\rangle_{\rm model}$} \\
    \colhead{Star}&  \colhead{(Mx)} &  \colhead{(kG)} & \colhead{(kG)}
    }
    \startdata
    UV Psc A   &  $25.77\pm0.45$  &  $0.79^{+1.43}_{-0.51}$ & $2.0$ \\
    UV Psc B   &     \nodata     &  $1.39^{+2.53}_{-0.90}$ & $4.6$ \\
    YY Gem A   &  $25.16\pm0.45$  &  $0.62^{+1.13}_{-0.40}$ & $4.3$ \\
    YY Gem B   &     \nodata    &\nodata&\nodata \\
    CU Cnc A   &  $24.85\pm0.45$  &  $0.62^{+1.13}_{-0.40}$ & $4.0$ \\
    CU Cnc B   &     \nodata    &  $0.76^{+1.38}_{-0.49}$ & $3.6$
    \enddata
    \label{tab:rad_bfield}
\end{deluxetable}

We have re-derived the empirical scaling relation with this expanded data
set (see the Appendix). The updated relation is shown in 
Figure \ref{fig:lx_mx_rad}, where we find that
\begin{equation}
    \log_{10}\Phi = \left(11.86\pm0.68\right) + \left(0.459\pm0.018\right)
        \log_{10}L_{x}.
    \label{eq:ols}
\end{equation}
Note that we have reversed the axes from how the relation was originally 
presented by \citet{Fisher1998}. This is because we are interested in
predicting $\Phi$ from a measure of $L_x$ instead of establishing a casual
relationship between the two quantities.

\subsection{Surface Magnetic Field Strengths}
\label{sec:surf_field_str}

X-ray properties of the three DEBs analyzed in this study are determined 
using X-ray data from the {\it ROSAT} All-Sky Survey Bright Source Catalogue 
\citep{Voges1999}. {\it ROSAT} count rates ($X_{\rm cr}$) and hardness ratios 
(HRs) are given in Table \ref{tab:rad_rosat}. We convert to X-ray fluxes 
using the calibration of \citet{Schmitt1995}. The conversion to X-ray fluxes 
is complicated by the fact that {\it ROSAT} has relatively poor spatial resolution, 
meaning any nearby companions to these DEBs may also be contributing to 
the total $X_{\rm cr}$. 

The X-ray flux quoted in Table \ref{tab:rad_rosat} is therefore calculated
as the estimated flux per star, determined by dividing the total flux 
by the number of stars thought to be contributing to $X_{\rm cr}$.
This is not a problem for UV Psc, which appears isolated. YY Gem and CU 
Cnc, on the other hand, have known, nearby companions. A search of the 
{\it ROSAT} Bright Source Catalogue for Castor A and B yield the same data as 
is found when searching for YY Gem, indicating that {\it ROSAT} cannot spatially 
resolve these three systems. Castor A and B are both binaries, thought
to have M-dwarfs companions, meaning that {\it ROSAT} is detecting X-ray emission
from up to six sources. Similarly, CU Cnc has a $12^{\prime\prime}$ companion, CV Cnc, 
another M-dwarf binary system. Both systems are likely contributing 
to the $X_{\rm cr}$ listed in the {\it ROSAT} Bright Source Catalogue. 

Distances to the systems are calculated using updated \emph{Hipparcos} 
parallaxes \citep{hip07}, except for YY Gem, for which no parallax is 
available. Instead, we adopt a distance of $13\pm2$ pc, which has been 
estimated from earlier investigations \citep{CB1995,Lopezm2007,FC12}.
The distances allow for the calculation of the total X-ray luminosity 
per star, which we quote in Table \ref{tab:rad_rosat}.

Our estimates for the DEB surface magnetic fluxes, surface magnetic field strengths, 
and their associated errors are given in Table \ref{tab:rad_bfield}. 
We also include the surface magnetic field strengths required by our models
for comparison. The X-ray data reveal that the magnetic field strengths 
required by our models are probably too strong. We demonstrate this 
visually by the dark blue diamond symbols in Figure \ref{fig:lx_mx_rad}.
Since the magnetic models reproduce the observed stellar radii, the surface
area of the model is equal to the observed stellar surface area. Thus,
the larger magnetic fluxes observed from the models suggest the magnetic
field strengths are too strong.

\begin{figure*}[t]
    \plottwo{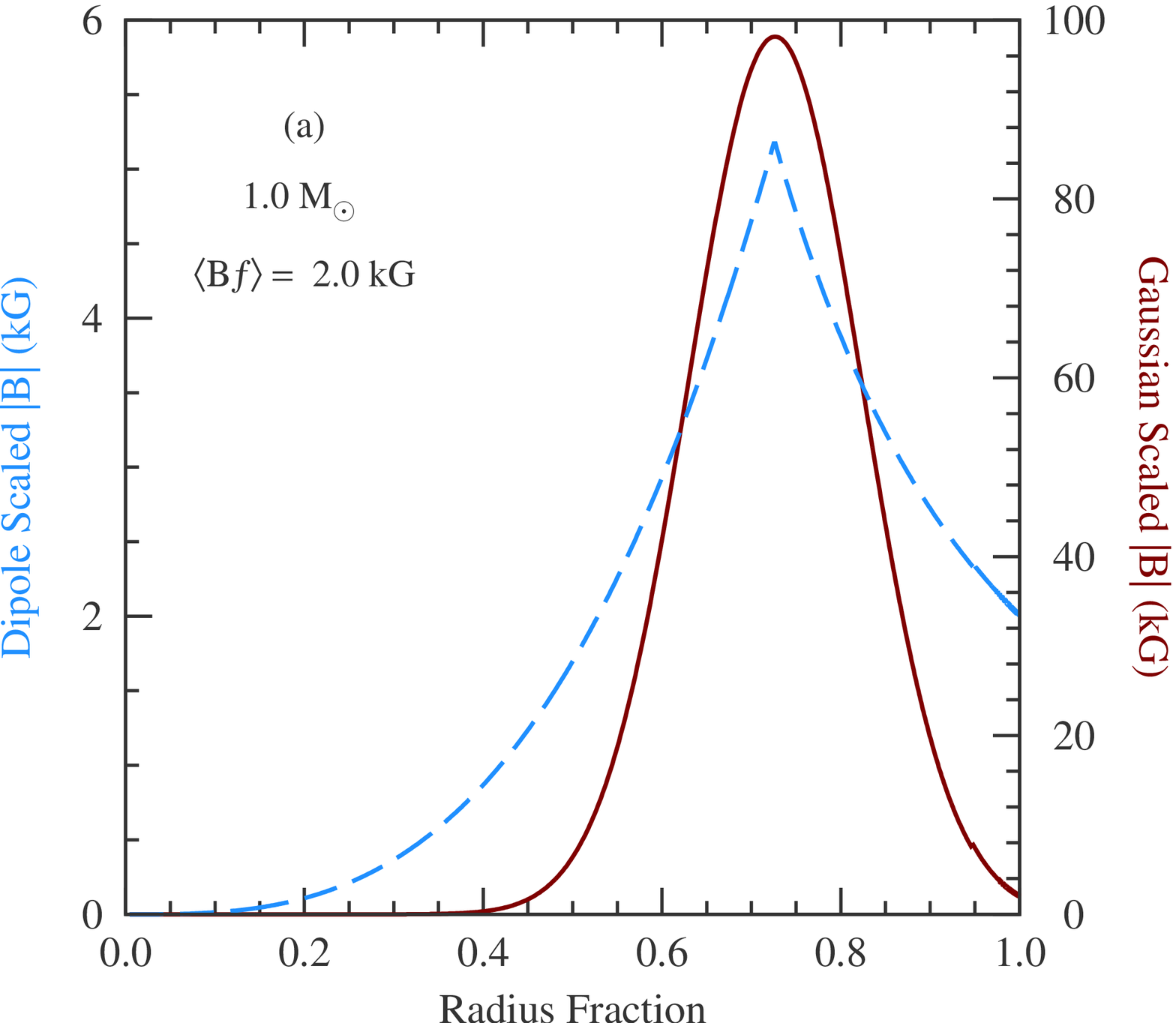}{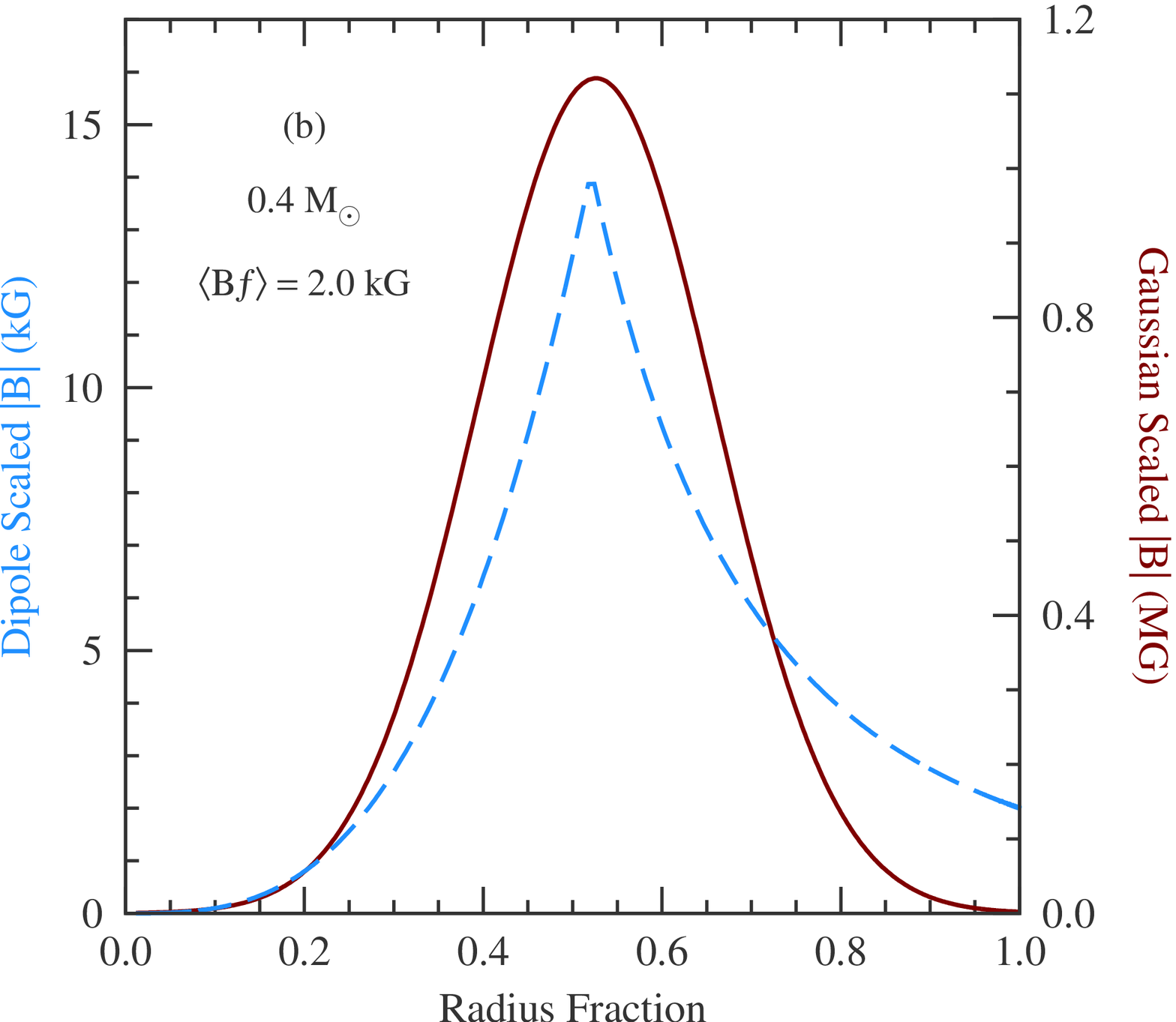}
    \caption{Comparison of the 
        dipole (light blue, dashed) and Gaussian (maroon, solid) radial 
        profiles for the interior magnetic field strength. (a) A $1.0\msun$ 
        model. (b) A $0.4\msun$ model. Note that the two field profiles
        are plotted on different scales with the dipole on the left $y$-axis
        and the Gaussian on the right $y$-axis.
        \vspace{0.5\baselineskip}\\
        (A color version of this figure is available in the online journal.)
        }
    \label{fig:rad_prof}
\end{figure*}

The data suggest that the only realistic magnetic field strengths may be 
those of UV Psc. However, the data points for UV Psc lie outside the domain 
of the empirical data set. Whether this extrapolation is valid remains 
unclear. Even if we assume the extrapolation is
valid, UV Psc B lies just above the $1\sigma$ error bounds of the linear 
correlation. Conversely, UV Psc A lies just inside the boundary. We must 
therefore be cautious with our interpretation of the accuracy of our models 
for UV Psc. This is further reinforced by noting that only two G-dwarfs 
(the spectral type of UV Psc A) were included in the data set and are 
therefore under represented. Direct measurements of G-dwarf surface magnetic 
field strengths show they are around $500$ G \citep{Saar1990}. This is 
consistent with the surface magnetic field strength derived for UV Psc A 
from the X-ray luminosity relation, but inconsistent with our model 
predictions.

What is clear from Figure \ref{fig:lx_mx_rad} and Table \ref{tab:rad_bfield}
is that the magnetic field strengths required by the models are too strong
for the stars in YY Gem and CU Cnc. The models of YY Gem need a magnetic 
field strength that are a factor of six too strong. Similarly large magnetic
fields are needed by the models for the stars of CU Cnc. Taking a more 
qualitative approach, the magnetic field strengths required by our models
for YY Gem and CU Cnc appear within the realm of possibility. Numerous studies
of M-dwarfs find that surface magnetic strengths of a few kilogauss are
quite common \citep{Saar1996,Saar2001,Reiners2007,Reiners2009}. 
However, these DEBs would then have noticeably stronger magnetic fields 
for their observed X-ray luminosity than the rest of M-dwarf population.
It appears that our approach to modeling magnetic fields in low-mass stars
may be incomplete.

\begin{deluxetable}{l c c}[t]
    \tablewidth{0.70\columnwidth}
    \tablecaption{Model Peak Interior Magnetic Field Strengths.}
    \tablehead{
    \colhead{DEB} & \colhead{$|B|_{\rm dipole}$} & \colhead{$|B|_{\rm Gaussian}$} \\
    \colhead{Star} & \colhead{(kG)} & \colhead{(kG)}
    }
    \startdata
    UV Psc A  &      4      &     40   \\
    UV Psc B  &     12      &    400   \\
    YY Gem A  &     13      &    500   \\
    YY Gem B  &   \nodata   &  \nodata \\
    CU Cnc A  &     21      &    1500  \\
    CU Cnc B  &     21      &    2000 
    \enddata
    \label{tab:int_fs}
\end{deluxetable}

\subsection{Interior Magnetic Field Strengths}
\label{sec:mag_rad_inter}

Assessing the validity of our predicted interior magnetic field strengths
is inherently more difficult. There is currently no reliable method for 
measuring the magnetic field strengths present inside stars. Therefore,
we elect to compare the theoretical magnetic field strengths required by 
our models with those predicted by 3D MHD models \citep[see, e.g.,][]{
Brandenburg2005}. 

Table \ref{tab:int_fs} presents the peak interior magnetic field strengths 
for the models presented above for both the dipole and the Gaussian field 
profiles (see below). The peak magnetic field strengths in Table \ref{tab:int_fs} 
are pre-defined to be at the base of the stellar convection zone \citep{FC12b}. 
They are all $\sim10^3$ G to $\sim10^4$ G. By comparison, 3D MHD
models routinely find peak magnetic field strengths of a few times $10^3$ 
G (consistent with equipartition estimates) to $\sim10^5$ G for 
solar-like stars \citep[see review by][]{Brandenburg2005}. In the immediate 
context, ``solar-like'' is loosely taken to mean stars with a radiative core 
and a solar-like rotation profile. The peak magnetic field strengths of 
our models appear consistent with those predicted from 3D MHD models.
Furthermore, while helioseismic investigations have yet to reveal 
the interior magnetic field profile for our Sun, initial indications suggest 
the peak magnetic field strength is below $300$ kG \citep{Antia2000}. Estimates
place the strengths in the vicinity of several tens of kG \citep{Antia2003}.

\subsection{Reducing the Magnetic Field Strengths}

Given the results that the model surface field strengths are likely too 
strong, we seek to reformulate our magnetic perturbation. We first consider 
some of the assumptions used to formulate our magnetic models presented 
in \citet{FC12b} and used in Section \ref{sec:ind_deb}. We have identified three
key assumptions: (1) our prescribed magnetic field radial profile, (2) that
convective elements are spherical and that the plasma obeys the equations of 
ideal magnetohydrodynamics (MHD), and (3) that the dynamo is driven completely 
by rotation.  

\subsubsection{Magnetic Field Radial Profile}
\label{sec:gauss_prof}

In the models presented up to this point, we have followed the formulation
described in \citet{FC12b}. There, we arbitrarily assumed that the magnetic 
field strength radial profile was a dipole configuration falling off as 
$r^3$ from a pre-defined peak field strength location. The location of the
peak magnetic field strength was taken to be at the base of the convective 
envelope, nominally the stellar tachocline. Note that in fully convective
objects, the peak magnetic field strength is located at $0.15 R_{\star}$
\citep{FC12b}. It may be possible to reduce 
the required surface magnetic field strengths if we instead use steeper 
radial profile. This would produce a stronger magnetic field at the 
tachocline for a given surface magnetic field strength.

To this end, we define a Gaussian radial profile. The tachocline is still
the location of the peak magnetic field strength, now defined to be the
peak of a Gaussian distribution. The radial profile is then defined by
\begin{equation}
    B(r) = B(R_{\rm src}) \exp\left[ -\frac{1}{2}\left( \frac{R_{\rm src} 
       - r}{\sigma_{g}}\right)^2 \right]
\end{equation}
where $\sigma_g$ is the width of the Gaussian and $R_{\rm src}$ is the 
radial location of the tachocline. The adopted width of 
the Gaussian is arbitrary. In most applications, it would be reasonable 
to set the width of the Gaussian to a constant value. For instance, a 
value of $\sigma_{g} = 0.2$ was favored by \citet{LS95}. However, since 
we will be considering stars with varying convection zone depths 
\citep[not only the Sun as in][]{LS95}, we felt a single value would 
not be appropriate. Instead, we define $\sigma_g$ as a function of the
convection zone depth. This allows us to localize the magnetic field in
thin convection zones and distribute the magnetic field in fully convective
stars, thereby maintaining seemingly realistic values for the peak magnetic
field strength (see Table \ref{tab:int_fs}). The width of the Gaussian was fixed 
to $\sigma_{g} = 0.2$ in fully convective objects and $\sigma_{g} = 0.1$ 
in the Sun. Therefore, we can define
\begin{equation}
    \sigma_{g} = 0.2264 - 0.1776\left(R_{\rm src}/R_{\star}\right).
\end{equation}
A direct comparison of the shape of the magnetic field profiles used
in this study is given in Figure \ref{fig:rad_prof}. Two masses are shown
to make clear the variable width of the Gaussian.

\begin{figure}
    \plotone{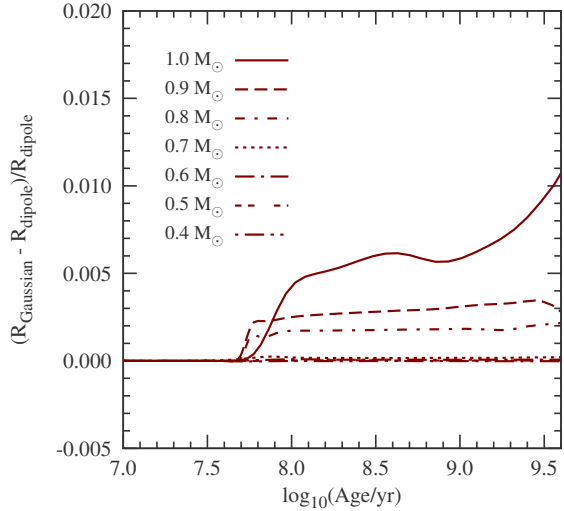}
    \caption{Effect on stellar radius predictions caused by using
        a Gaussian radial profile instead of a dipole profile. Shown is 
        the relative radius difference as a function of age for a series 
        of stellar models. All of the models have equivalent surface magnetic 
        field strengths (2.0 kG) but different magnetic field radial profiles.
        \vspace{0.5\baselineskip}\\
        (A color version of this figure is available in the online journal.)
        }
    \label{fig:gau_dip}
\end{figure}

Figure \ref{fig:gau_dip} illustrates the influence of using the Gaussian 
radial profile. We plot the relative difference in the radius evolution 
between the Gaussian and dipole radial profiles for a series of stellar 
masses. All of the models have an equivalent surface magnetic field strength 
of 2.0 kG and a solar metallicity. Despite the increased magnetic field 
strength at the tachocline (see Table \ref{tab:int_fs} and Figure 
\ref{fig:gau_dip}), using a Gaussian radial profile instead of a dipole 
profile has a largely sub-1\% effect on model radius predictions. Therefore, 
we find no compelling reason to alter our default field strength profile.

Two additional comments on the results presented in Figure \ref{fig:gau_dip}.
First, the two radial profiles produce similar results despite the peak
magnetic field strengths differing by about an order of magnitude. It would 
seem that the deep interior field strength is relatively unimportant in 
stars with radiative cores. Instead, the surface magnetic field strength 
appears to be primarily responsible for driving the radius inflation. This
is consistent with the previous study by \citet{Spruit1986}. 

\begin{figure}[t]
    \plotone{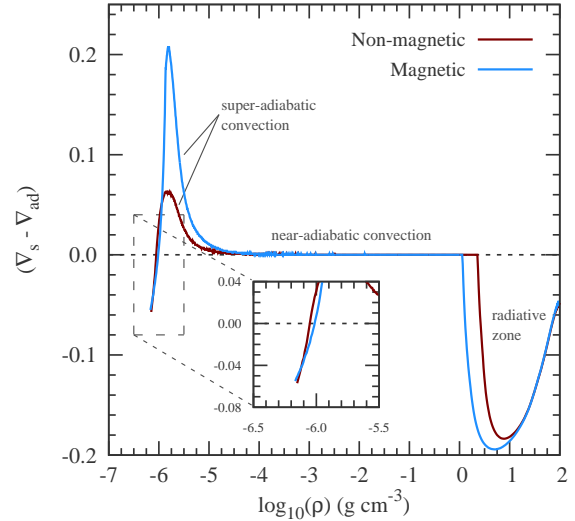}
    \caption{Difference between the plasma temperature gradient,
        $\tgrad$, and the adiabatic temperature gradient, $\dela$, as a
        function of the logarithmic plasma density for a $M=0.6\msun$ star. 
        We show this for two
        models: a non-magnetic model (maroon, solid line) and a magnetic 
        model (light-blue, solid). The zero point is marked by a gray
        dashed line, dividing locations where convection (positive) or
        radiation (negative) is the dominant flux transport mechanism. 
        The inset zooms in on the near-surface region where radiation 
        becomes the dominant flux transport mechanism.
        \vspace{0.5\baselineskip}\\
        (A color version of this figure is available in the online journal.)
        }
    \label{fig:del_dela}
\end{figure}

To understand why, we plot the difference between the temperature gradient
of the ambient plasma, $\tgrad$, and the adiabatic gradient, $\dela$, as 
a function of density in Figure \ref{fig:del_dela}. Consider the non-magnetic 
model. When the 
line dips below the zero point, radiation carries all of the excess energy. 
Near the stellar surface, where the density, $\rho$, is small, $\tgrad$ is noticeably 
super-adiabatic, indicative of inefficient convective energy transport. 
Deeper in the star, where $-5 < \log_{10}{\rho} < 0$, the temperature
gradient is super-adiabatic, but only slightly ($\tgrad - \dela < 10^{-8}$). 
This suggests convective energy transport is highly efficient. 

Our magnetic field implementation modifies $\dela$ by a factor proportional 
to $\nu\delx$, where
\begin{equation}
    \nu\delx = \frac{P_{\rm mag}}{P_{\rm gas} + P_{\rm mag}}
        \left(\frac{d\ln\chi}{d\ln P}\right).
    \label{eq:nu_delx}
\end{equation} 
In Equation (\ref{eq:nu_delx}), $\nu = -(\partial\ln\rho/\partial\ln\chi)$
at constant $P$ and $T$, a magnetic compression coefficient, and $\delx$
is the gradient of the magnetic energy per unit mass, $\chi\ (= B^2/8\pi\rho)$, with respect
to the total pressure, $P$ (also see Equation (\ref{eq:fc12_nu})). Deep in the
stellar interior, $\delx \sim 0.1$, and $\nu \sim 10^{-8}$ -- $10^{-9}$. 
This is not sufficient to inhibit convection deep within the star. The exact 
situation can be more complicated and is discussed further in Section \ref{sec:disc}. 
Near the surface, however, any inhibition of convection causes a steepening 
of $\tgrad$, forcing radiation to carry more energy. We then expect
to see a growth of the region near the stellar surface where radiation 
carries all of the flux, as is seen in Figure \ref{fig:del_dela}.
The structural changes caused by the steep temperature gradient near the
surface are enough to reconcile the models with the observations before 
the deep interior magnetic field strength becomes appreciable in magnitude
so as to inhibit convection. Therefore, the outward movement of the convection 
zone boundary occurs largely as a response to changes near the stellar 
surface.

Next, we see in Figure \ref{fig:gau_dip} that the $1.0\msun$ track
displays a sharp upturn near 1 Gyr. The model is too young 
to be undergoing rapid evolutionary changes. Instead, we believe this is 
related to the physical properties of the star near the boundary of the 
convection zone. As we just discussed, the magnetic field strengths for 
both profiles are typically too weak near the tachocline to affect stellar 
structure. However, this is not true for a $1.0\msun$ model. The 
convection zone is thin and convection is generally super-adiabatic 
throughout. This results in the Gaussian radial profile having a stronger 
effect on the properties of convection. We now also see why the Gaussian 
radial profile has a weaker influence on the stellar radius as we decrease 
the stellar mass. Convection in the deep interior becomes more adiabatic 
(re: efficient) at lower masses.

\begin{figure}[t]
    \plotone{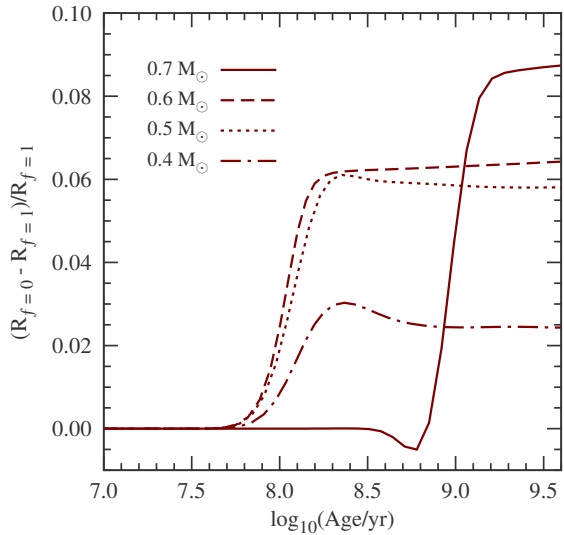}
    \caption{Influence of the parameter $f$ on model radius predictions.
         Plotted is the relative difference in radius evolution between
         magnetic models computed with $f=1$ and $f=0$. All models have
         a prescribed surface magnetic field strength of $2.0$ kG.
         \vspace{0.5\baselineskip}\\
        (A color version of this figure is available in the online journal.)
        }
    \label{fig:f_par}
\end{figure}

\subsubsection{The Free Parameter, $f$}

The technique advanced by \citet{LS95} introduces a number of free
parameters that govern the interaction between an imposed magnetic 
field and the stellar plasma. One of these parameters, $\gamma$, governs 
the amount of pressure exerted on the plasma by a magnetic field and 
was investigated in \citet{FC12b}. Another parameter, $f$, controls the 
relationship between the magnetic energy gradient of a convecting 
element and that of the surrounding plasma. Physically, $f$ is related to 
the geometry of the convecting element and the conductivity of the plasma. 
For instance, assuming that convecting elements are spherical and that 
the plasma is infinitely conducting, we found that $f \approx 1$ \citep{FC12b}. 
Although stellar interiors may be considered, for all practical purposes, infinitely
conducting, this is not necessarily the case in the stellar envelope.

Changing $f$ so that it is no longer fixed to 1 introduces an additional term
in the convective buoyancy equation. This term can either increase or decrease
the buoyancy force felt by a convecting element, depending on the sign of the
plasma magnetic energy gradient and the precise value of $f$ 
\citep[see Equation (57) in][]{FC12b}. In our studies, the magnetic energy 
gradient is decreasing toward the model center throughout the 
stellar envelope. Thus, decreasing $f$ from 1 slows the buoyant motion 
of a convective element thereby reducing the convective flux. 

Instead of attempting to physically motivate a value for $f$, we take the approach
of testing an extremal value. By setting $f=0$ we maximize the Lorentz force term 
mentioned above, given our radial profile. Results of changing the value of $f$ 
from $f=1$ to $f=0$ are shown in 
Figure \ref{fig:f_par}. As with Figure \ref{fig:gau_dip}, we plot the relative 
difference in the radius evolution for a series of mass tracks. All
of the models were calculated with a $2.0$ kG magnetic field and a dipole
field profile (see the previous section). The largest mass model shown 
is $0.7\msun$ as models with larger masses failed to converge for 
a $2.0$ kG magnetic field with $f = 0$. It is evident from Figure \ref{fig:f_par} 
that lowering the value of $f$ can have a significant impact on the model 
radius evolution. Differences between 2\% and 9\% were observed, depending 
on the mass of the model. Therefore, it is worthwhile to study the effect 
of setting $f=0$ on the individual DEB stars discussed above.

Setting $f = 0$ reduced the magnetic field strengths required to reconcile
model radii by about half. The model of YY Gem, for instance, only requires
a surface magnetic field strength of 2.5 kG compared to the 4.3 kG required
with $f = 1$. However, the reduced surface magnetic field strengths are 
still too strong compared with values estimated from X-ray emission. Models
of UV Psc are in better agreement with the X-ray predicted values, but we 
have already given reasons to be skeptical of these predictions.

Whether or not we have reason to believe that $f$ should be zero is a more
difficult question. Geometric considerations about convecting elements have
long been debated in standard MLT, but they ultimately have little impact on 
model predictions \citep{Henyey1965}. However, when considering magnetic
MLT, convective element geometry may come to matter. Quantifying the effects
of plasma conductivity on $f$ is far more difficult. It is safe to assume that the
stellar plasma is highly conducting throughout a majority of the stellar interior,
where temperatures exceed $10^4$\,K and hydrogen is completely ionized. 
Near the surface, free electrons from hydrogen ionization are no longer 
available, meaning
conductivity is decreased, but not necessarily zero. Free electrons from other 
species (e.g., Ca and Na) keep the plasma at least partially conducting. 
Physically motivating an expression for $f$ in terms of the plasma conductivity 
and convective bubble geometry would be a worthwhile endeavor. We plan
to investigate this in a future study. Ultimately, we have shown that the value
of $f$, while of consequence, does provide enough leverage to bring our
models into agreement with the observations using realistic surface magnetic
field strengths.

\subsubsection{Dynamo Energy Source}
\label{sec:des}

The final assumption that we identified was that the magnetic field is 
``driven'' completely by rotation. By doing so, we have allowed
for the unintended consequence that the magnetic field strength---both 
the surface and the interior---can grow without limit.\footnote{We note 
that there is a computational limit whereby too strong of a magnetic field
will prevent the model from converging to a solution. This occurs slightly
after the magnetic pressure exceeds the gas pressure near the photosphere.} 
There is no natural limit imposed upon the field strengths as the 
mechanism from which the field is drawing energy (i.e., rotation) is 
neglected entirely.

Limiting the magnetic field strengths, however, does not alter the ease 
by which a magnetic field can induce stellar inflation. What it can do is 
validate or invalidate the required surface magnetic field strengths. However, 
if we reconsider the physical source of the magnetic field, it can lead 
to new methods of including magnetic fields in stellar evolution codes.
There are questions as to whether the solar magnetic field is generated at 
the tachocline by the strong shear induced by rotation \citep[i.e., the 
standard Parker model;][]{Parker1955} or if it is primarily generated 
within the solar convection zone by turbulent convection without explicit 
need for a tachocline \citep{Brandenburg2005,Brown2010}.

Generation of a magnetic field from turbulent convection would suppress 
convective velocity fluctuations thereby reducing the total heat flux 
transported by convection. Early three-dimensional magneto-convection 
simulations support this assessment \citep{Stein1992}. Given that thick 
convective envelopes are a ubiquitous feature of low-mass stars, it
is not unreasonable to posit that suppressing the total heat flux transported
by convection would strongly impact stellar structure.

Assuming that the magnetic field sources its energy directly from the 
kinetic energy of turbulent convection has important consequences for a 
magnetic theory of convection. Consider a single convecting bubble. If the
bubble is traveling with some velocity, $u_{\rm conv, \, 0}$, it has a 
kinetic energy equal to $\rho u_{\rm conv, \, 0}^2/2$, where $\rho$ is the 
mass density of the plasma. Now, suppose some of that kinetic energy is 
used to generate the local magnetic field. By conservation of energy, we 
must have that
\begin{equation}
\frac{1}{2}\rho u_{\rm conv, \, 0}^2 = \frac{1}{2}\rho\uconv^2 + \frac{B^2}{8\pi},
\end{equation}
where $\uconv$ is the convective velocity after generation of a magnetic 
field and the final term is the magnetic energy. We have implicitly 
assumed that a characteristic convecting bubble is responsible for generating 
the magnetic field only in its vicinity. This local treatment is a purely
phenomenological approach and gives an upper limit to the effects of
a turbulent dynamo, but provides a reasonable zeroth-order approximation
within the framework of MLT.

The result is that the characteristic convective bubble will have a lower
velocity in the presence of a magnetic field,
\begin{equation}
\uconv^2 = u_{\rm conv, \, 0}^2 - u_A^2.
\label{eq:u_conv_t}
\end{equation}
The last term on the right hand side is equal the square of the local 
Alfv\'{e}n velocity,
\begin{equation}
u_A^2 = \frac{B^2}{4\pi\rho}.
\end{equation}
The diversion of energy into the magnetic field can significantly
reduce the total convective flux, as the convective flux is proportional
to $\uconv^3$. This reduction in convective flux forces the radiative
temperature gradient to grow steeper by an amount proportional to $u_A^2$.

To include these effects, we re-derived the equations of MLT
given in \citet{FC12b}.  The convective velocity and convective flux were
modified to account for the loss of energy to the magnetic field. Unfortunately,
we were unable to find convergence when numerically solving the system of 
equations. Closer inspection of the coefficients in our new quartic equation
\citep[see Equation (65) in][]{FC12b} revealed that there was no real root
in the outer portions of the stellar envelope. 
The low density present in the envelope drives up $u_A^2$, causing the 
range of our final equation to lie above zero for all real values of the
convective velocity. We also attempted to modify the ``non-adiabaticity'' 
equation \citep[Equation (58) in][]{FC12b} so as to model the transformation 
of turbulent kinetic energy into magnetic energy as an effective heat loss. 
However, we were again unable to obtain convergence for the resulting
quartic equation. We are continuing the investigate this issue.

Instead, we opted to first solve the MLT equations as normal and then
modify the results to mimic the conversion of turbulent kinetic 
energy to magnetic energy. The convective velocity is reduced by the 
Alfv\'{e}n velocity, as in Equation (\ref{eq:u_conv_t}). While energy is 
transferred to the magnetic field, we assume that the energy is still 
confined to the local region under consideration, thereby reducing the total 
flux of energy across the region. Assuming that energy flux is conserved,
radiation must attempt to carry additional energy flux. This leads to an
increase in the local temperature gradient of the ambient plasma that is
proportional to the total energy removed from the convecting element. 
Specifically,
\begin{equation}
\Delta(\tgrad - \dele) \propto \frac{u_A^2}{C},
\end{equation}
where $C = g \amlt^2 H_P \delta/8$ is the characteristic squared velocity 
of an unimpeded convecting bubble over a pressure scale height ($H_P$). 
In the definition of $C$, we find the local gravitational acceleration $g$,
the convective mixing length parameter $\amlt$, and the coefficient of 
thermal expansion $\delta = -(\partial\ln\rho/\partial\ln T)_{P,\, \chi}$.
Increasing the temperature excess increases the buoyancy of a convective
element, increasing its velocity. However, for now, we concern 
ourselves only with the simple first-order approximation and neglect the 
immediate feedback on the convective velocity. 

\begin{figure}[t]
    \plotone{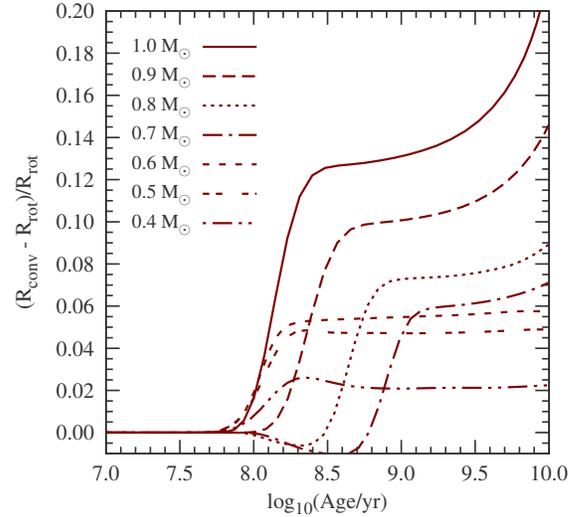}
    \caption{Effect on model radius predictions when the dynamo source 
        is turbulent convection instead of purely rotation. Shown is 
        the relative radius difference as a function of age for a series 
        of stellar models. All of the models have an equivalent surface 
        magnetic field strengths of $0.5$ kG.
        \vspace{0.5\baselineskip}\\
        (A color version of this figure is available in the online journal.)
        }
    \label{fig:mlt_new_ab}
\end{figure}

We again compute a series of models using this new modified MLT and
compare the relative differences in radius evolution to our original formulation. 
These tracks are
shown in Figure \ref{fig:mlt_new_ab}. We assume a $500$ G surface magnetic
field strength with a dipole radial profile and $f=1$. Stabilization of convection,
or a modified Schwarzschild criterion, is neglected, leaving only the reduction
of convective flux to influence the convective properties. The $0.7\msun$,
$0.8\msun$, and $0.9\msun$ tracks in Figure \ref{fig:mlt_new_ab} show
slightly different characteristics. Models assuming a turbulent dynamo
readjusted to the presence of the magnetic field at an older age than did
the models with the rotational dynamo. Despite appearances, all perturbations 
were introduced at an age of $\tau_{\rm age} = 0.1$ Gyr. It is apparent
from Figure \ref{fig:mlt_new_ab} that this new method of handling magneto-convection
can significantly reduce the surface magnetic field strength required to
inflate model radii. 

We recomputed models for UV Psc, YY Gem, and CU Cnc using the best fit 
metallicities found in Section \ref{sec:ind_deb}. We were able to
achieve model convergence for the stars in YY Gem and CU Cnc. Models of
the stars in UV Psc with surface magnetic fields greater than $650$ G, 
required to correct the model radii, did not converge. We also recomputed
models for \object{EF Aquarii} \citep{vos12,FC12b}, but ran into the same 
convergence issues. We are continuing to investigate this issue. The resulting 
magnetic field strengths required for YY Gem and CU Cnc A and B were 
$0.7$ kG, $0.7$ kG, and $0.8$ kG, respectively. 

The surface magnetic field strengths derived from sourcing the magnetic field 
energy from convection are nearly identical to the value estimated from 
X-ray emission.
This agreement lends credence to our latest approach, at least for lower
mass models. We found it difficult to attain model convergence using this 
approach for models with masses greater than about $0.75\msun$. Stronger 
magnetic fields were required than could be achieved with the present 
model configuration. This was a consequence of the Alfv\'{e}n velocity 
exceeding the convective velocity at some point within the convective
envelope. Below that approximate mass limit the turbulent dynamo approach 
can inflate stellar radii with relative ease compared to our original 
formulation. Whether this is indicative of the actual dynamo processes 
acting in each star is not clear. It is thought that the dynamo mechanism
begins shifting from a rotationally driven dynamo to a turbulent dynamo 
somewhere around $0.6\msun$ \citep[early-M; pg.\ 228 in][and references
therein]{Reid2005}, which would suggest YY Gem has a predominantly
rotational dynamo. We explore possible tests to delineate between the 
two processes using stellar models in the next section.

\section{Discussion}
\label{sec:disc}

\subsection{Interior Structure}

Section \ref{sec:des} presents a second approach to modeling 
the effects of magnetic fields on thermal convection. Instead of acting 
to stabilize convection through the modification of the Schwarzschild criterion, 
the method acts to reduce the efficiency of convection. Ultimately, the 
two approaches represent different physical mechanisms by which a star 
may produce its magnetic field. 

A dynamo that sources its energy primarily from rotation will act to stabilize 
convection and will not drain the energy from convective elements. On the 
other hand, a turbulent dynamo will tend to exhaust the kinetic energy available 
from convection. The magnetic field created by a turbulent dynamo may also act 
to stabilize convection, but this effect is secondary in our models. The magnetic
field strengths required to inflate a star following the turbulent dynamo 
approach are considerably weaker than those required when assuming a rotationally
driven dynamo. Given the two different mechanisms that may be producing 
stellar magnetic fields, it is worth exploring whether or not the two 
approach produce any discernible differences between two otherwise identical
stars. We have found the surface magnetic field strengths will be different,
but is the stellar interior structure different between the two? 

\begin{figure}[t]
    \plotone{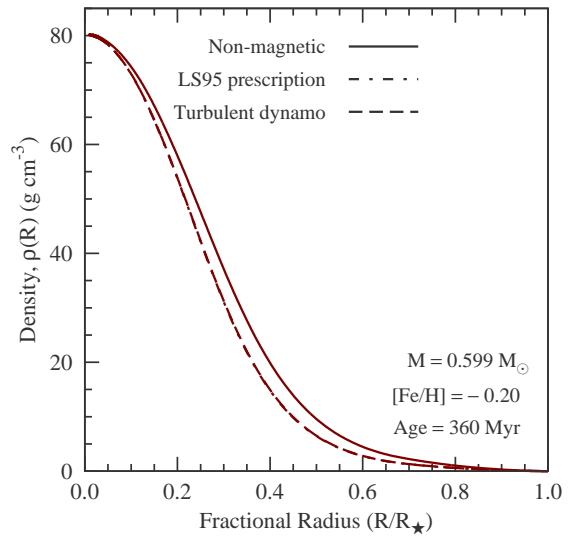}
    \caption{Interior density profile for a $M = 0.599\msun$
        model with and without the presence of a magnetic field. The magnetic 
        models were computed with a surface magnetic field strength strong 
        enough to reconcile the model radius with the observed radius of 
        YY Gem ($4.3$ kG and $0.7$ kG for the \citet{FC12b} prescription
        and turbulent dynamo approach, respectively). Note that the two 
        lines for the magnetic models directly on top of one another.
        \vspace{0.5\baselineskip}\\
        (A color version of this figure is available in the online journal.)
        }
    \label{fig:int_prof}
\end{figure}

Figure \ref{fig:int_prof} shows the run of density within three different model
interiors: one standard, non-magnetic model and two magnetic models with
different dynamo methods.  Each model was computed with a mass of $0.599 \msun$, 
a metallicity of $-0.20$ dex, and evolved to an age of 360 Myr---the properties 
of the YY Gem stars. The non-magnetic model has a radius smaller than the 
actual radius of the YY Gem (here considered a single star). Surface magnetic
field strengths used in the magnetic models were those required to reconcile
model radii with observations ($4.3$ kG and $0.7$ kG for the \citealt{FC12b} 
prescription and turbulent dynamo, respectively).

Overall, introducing a magnetic field causes the density profile to steepen
within the models with little to no change in the central density. As can be
seen in Figure \ref{fig:int_prof}, the magneto-convection methods produce
identical results (the two lines are lying on top of one another) despite having
different magnetic field strengths throughout. Only near the surface of the
model (outer  0.2\% by radius) do the two methods produce noticeable 
differences. We compared the sound speed profiles of the three models, 
as well, and found similar results. The two magnetic models were characterized
by slower sound speed throughout compared to the standard model (as 
expected since they show lower densities). Differences between the two
magnetic models were again noted in the outermost layers, where the sound
speed of the turbulent dynamo model followed the sound speed of the non-magnetic
model before deviating and tracking the rotational dynamo model throughout 
the rest of the model interior. Even if one could perform seismology on stars
in this mass range, differences in the outermost layers would be unobservable.
Thus, for all practical purposes, the two magnetic models produce identical 
results.

Finally, we confirmed that the radius to the convection zone boundary was 
equivalent in the magnetic models. In the non-magnetic model, the convection 
zone boundary was located at $R/R_\star = 0.671$, but it receded to 
$R/R_\star = 0.691$ and $R/R_\star = 0.690$ for the magnetic models. 
Therefore, it appears that the only way to differentiate between the two proposed 
methods is to measure a star's surface magnetic field strength. 

\subsection{Comparison to Previous Work}
\label{sec:mag_rad_prev}

\begin{figure}
	\plotone{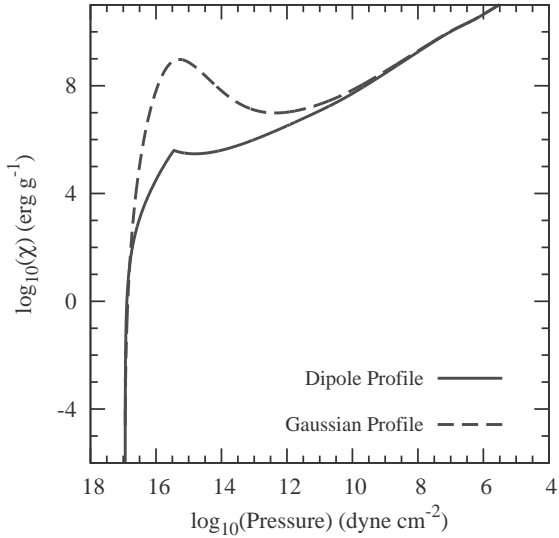}
	\caption{Variation of the magnetic energy per unit mass, $\chi$, vs.\ the 
	    total pressure shown for two magnetic field radial profiles: dipole
	    (solid line) and Gaussian (dashed line). A fixed mass ($M=0.5\msun$) 
	    and surface magnetic field strength ($\langle Bf\rangle = 2.0$\,kG) was used 
	    to generate each profile at a model age of 1\,Gyr.}
	 \label{fig:logxlogp}
\end{figure}

Similar methods for incorporating the effects of a magnetic field have 
been performed previously. These previous studies have separately looked
at magnetic stabilization of convection \citep{MM01} and the reduction
of convective efficiency \citep{Chabrier2007}. It is instructive to compare
the results presented in this paper with those earlier studies.

Qualitatively, our initial approach to modeling the magnetic interaction 
with convection \citep{FC12b} is very similar to the method favored by
\citet{MM01}. The variable $\nu$, a ``magnetic compression coefficient,'' appearing 
in our modified Schwarzschild criterion \citep[Equation (53) in][]{FC12b} 
is essentially equal to the \citet{MM01} convective inhibition parameter, 
$\deltamm$. Explicitly, 
\begin{equation}
    \nu = -\thermd{\ln\rho}{\ln\chi}{P}{T} =  
          \frac{B^2}{B^2 + 8\pi P_{\rm gas}}.
    \label{eq:fc12_nu}
\end{equation}
Comparing with the magnetic inhibition parameter presented by \citet{MM01},
\begin{equation}
    \deltamm = \frac{B^2}{B^2 + 4\pi\gamma P_{\rm gas}},
    \label{eq:mm01}
\end{equation}
where $\gamma$, the ratio of specific heats, is of order unity. Their formulation
is based on the work of \citet{GT66}. One significant difference is that 
our approach introduces a magnetic energy gradient, $\delx$, in the 
stability criterion. This gradient determines how the magnetic field, 
characterized by $\nu$, interacts with convection. The $\delx$ term appears 
in our equations because we allow the magnetic pressure gradient to 
influence the equilibrium density \citep[Equation (34) in][]{FC12b}. 
\citet{GT66} restricted themselves to simple field geometries in a medium
with uniform density. It is for this reason that we favor our approach 
and the inclusion of $\delx$ in the stability criterion.

Figure \ref{fig:logxlogp} shows how the magnetic energy varies as a 
function of pressure for our two radial profiles. Immediately it can be 
seen that the two profiles differ significantly near the base of the 
convection zone around $\log_{10}(P) \sim 15.5$. Throughout a large
portion of the convection zone, the gradient is dominated not by the
magnetic field radial profile, but by the density profile. Density increases
steeply with radius in the outer regions of the star. Since $\chi$ ($= B^2/8\pi\rho$)
is inversely
proportional to density, we see a decrease in $\chi$ in the outer layers.
A negative gradient in Figure \ref{fig:logxlogp} implies that the magnetic
field has a stabilizing effect on convection. Since the gradient in density
dominates, the effect of magneto-convection is rather independent of 
the radial profile (so long as the magnetic field is an increasing function
of depth from the surface).

Deeper in the star, where the two radial profiles begin to deviate, we see
that there is a change in slope for the Gaussian profile. In this region, the 
magnetic field strength gradient begins to dominate. Since the dipole
profile has a shallower slope through the interior the change in slope occurs
deeper in the star. A positive magnetic energy gradient in Figure \ref{fig:logxlogp}
has a destabilizing effect on convection---convection becomes more favorable.
However, since this change in slope occurs within an already convectively 
unstable region, the overall effect on stellar structure is minimal. One may
then postulate that the dipole profile, with it's almost continuous negative
magnetic energy gradient should induce larger changes on stellar structure.
But, to maintain a negative slope the magnetic field strength cannot rise too
rapidly, meaning the value of $\nu$ remains small. 

Given our model's dependence on $\delx$ in the stability criterion, we expect our
required magnetic field strengths to be about a factor of 2--5  larger than those 
required by \citet{MM01}.
Subsequent works incorporating the \citet{MM01} magnetic inhibition parameter
do not explicitly model any of the three systems we have presented \citep[e.g.,][]{
MM10,MM11,MacDonald2013}. However, drawing from the general conclusions 
of those works, it appears that they regularly require surface magnetic
field strengths of about 0.5\,kG to produce models consistent with observations.
This is approximately an order of magnitude less than what we have presented
in Section \ref{sec:ind_deb} using our original method \citep{FC12b}. Only part 
of this difference can be accounted for by our inclusion of $\delx$ in the stability 
criterion. The rest of the difference may result from our incorporation of the
magnetic field in the density equation of state. Additional tests must be carried
out to assess the various contributions to these differences.

The tactic used in Section \ref{sec:des} has attempted to quantify a 
reduction in convective efficiency. \citet{Chabrier2007} had previously 
explored this concept, although they did so by arbitrarily reducing $\amlt$.
Although reducing $\amlt$ can be considered conceptually different from
our approach in Section \ref{sec:des}, the two are remarkably similar. 

Convective efficiency can be defined using the framework of MLT by 
considering heat losses that are ``horizontal'' to the radial motion of 
a bubble \citep{bv58,Weiss2005}. The efficiency, $\Gamma$, can be expressed 
as
\begin{equation}
    \Gamma = \frac{c_P}{6ac}\frac{\kappa \rho^2 \uconv \amlt H_P}{T^3},
\end{equation}
where $c_P$ is the specific heat at constant pressure, $a$ is the radiation
constant, $c$ is the speed of light in a vacuum, and $\kappa$ is the opacity.
Since the convective efficiency is proportional to both $\uconv$ and $\amlt$,
we may compare reductions in $\uconv$ with the reductions in $\amlt$ used 
by \citet{Chabrier2007}. 

Consider a reduction in $\uconv$ caused by a magnetic field that has a 
strength that is some fraction, $\Lambda$, of the equipartition field 
strength, $B_{\rm eq} = (4\pi\rho\uconv^2)^{1/2}$. Then the ratio of the 
convective efficiency in the presence of a magnetic field to the non-magnetic 
efficiency is $\Gamma_{\rm mag}/\Gamma_{0} = (1 - \Lambda^2)^{1/2}$. Therefore, 
to achieve the same results by reducing $\amlt$, we need only multiply 
the solar $\amlt$ by $(1 - \Lambda^2)^{1/2}$. We then find that
\begin{equation}
    \Lambda = \left[1 - \left(\frac{\amlt}{\alpha_{\rm MLT,\, \odot}}\right)^2\right]^{1/2}.
    \label{eq:lambda}
\end{equation}
This assumes that the convective velocity is unaffected by changes in 
mixing length. Allowing for changes in convective velocity, the exponent 
on the mixing length term in Equation (\ref{eq:lambda}) can be larger. 
We now have a quantitative
description of the degeneracy between magnetic inhibition of convection
and reducing the convective mixing length \citep{Cox1981,MM10} in terms
of a fraction of the equipartition magnetic field strength. 

\begin{figure}[t]
    \plotone{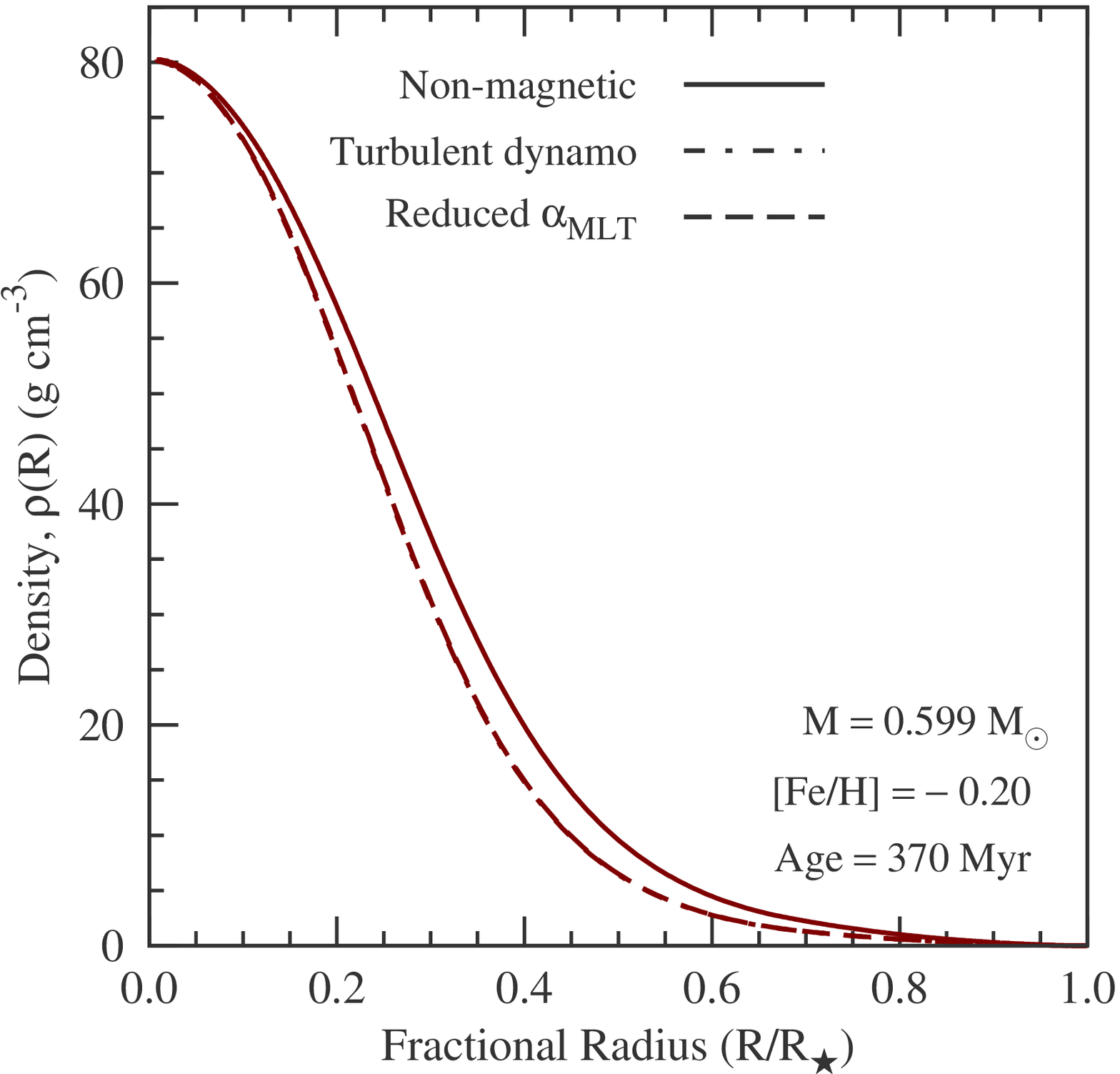}
    \caption{Similar to Figure \ref{fig:int_prof}, but comparing a reduced
        $\amlt$ model with a turbulent dynamo magnetic model. The turbulent
        dynamo model has a $0.7$ kG surface magnetic field and the reduced
        $\amlt$ model has $\amlt = 0.6$. Note that the magnetic model
        and reduced $\amlt$ model profiles lie on top of one another.
        \vspace{0.5\baselineskip}\\
        (A color version of this figure is available in the online journal.)
        }
    \label{fig:int_amlt_prof}
\end{figure}

\citet{Chabrier2007} 
considered multiple values for $\amlt$, including $\amlt = 0.1$
and $\amlt = 0.5$. These drastically reduced $\amlt$ values correspond 
to $\Lambda \sim 0.999$ and $\Lambda \sim 0.966$, respectively.
We find that our models do require such strong magnetic fields, at least in
the outer layers. The model of YY Gem in Section \ref{sec:des} has $\Lambda
\sim 1$ throughout a large portion of the outer envelope. $\Lambda$
decreases from $1$ at the photosphere (where $T=\teff$) to $0.3$ at the
point where we match the envelope to the interior. This is the result of
prescribing a magnetic field radial profile that is independent of $\Lambda$. 

To make the comparison more direct, we compare the interior structure of 
the magnetic model with that from a reduced $\amlt$ model. Our models 
required $\amlt = 0.60$ to reproduce the observed radius of YY Gem, 
corresponding to $\Lambda = 0.951$, for reference. Based on Figure 1(a) 
of \citet{Chabrier2007} it appears that they require $\amlt \sim 0.4$ 
to reproduce the properties of YY Gem.  We compare the density distribution
of our reduced $\amlt$ and magnetic model in Figure \ref{fig:int_amlt_prof}.
We find no significant difference. It is apparent that our magnetic model
produces results consistent with reduced $\amlt$ models.

\subsection{Star Spots}

Up to this point we have avoided any mention of specifically incorporating
effects due to star spots. Previous magnetic investigations have accounted 
for dark spots by reducing the total flux at the model photosphere \citep{Spruit1986,
Chabrier2007,Morales2010,MM11,MacDonald2013}. Reductions of photospheric 
flux were combined with the aforementioned magneto-convection techniques
to reconcile model radii with observations of DEBs \citep{Chabrier2007,
MM11}. However, as we have shown, our models do not explicitly require
spots to produce agreement with observations. 

This can be understood by considering that there is a degeneracy between
effects due to spots and effects due to magneto-convection \citep{MM10}. 
Generally, the total flux leaving the model photosphere is reduced. By including 
spots, magnetic field strengths required by the models to reproduce 
observations would be decreased. We elected to not include spots explicitly
for two reasons: (1) spots are the manifestation of inhibited, or suppressed,
convection, and (2) average surface magnetic field strengths are more easily
measured than spot properties.

Spots are surface blemishes caused by a reduction in convective energy
transport. Therefore, we are of the opinion that modeling effects of 
magnetic fields on convection is a more direct way to model effects from
spots. Convective properties required by our models represent the average 
global properties of convection, producing a convective flux that is 
between the increased convective flux in unspotted areas and the reduced
convective flux below spots. 

Possibly of greater importance, however, is that surface magnetic field 
strengths are directly observable. If effects of spots on stellar structure
are distinct from magneto-convection formulations, differences should 
appear between measured surface magnetic field strengths and those required
by stellar models. The problem with immediately including spots is that 
required spot properties are difficult to validate, especially for M-dwarfs.
Including spots in stellar models requires the specification of a free
parameter that describes a surface coverage of black ($T = 0$\,K) spots 
equivalent to a desired surface coverage of gray ($T > 0$\,K)
spots \citep{Spruit1986}. Measuring either the effective coverage of black
spots or the actual coverage of gray spots on real stars is a highly 
uncertain process. Therefore, it may be better to probe whether spots 
must be considered independently from magneto-convection by comparing pure
magneto-convection models to magnetic field observations to look for 
discrepancies between surface magnetic field strengths.

\subsection{Implications for Asteroseismology}

Figure \ref{fig:int_prof} showed that magnetic models have a lower density
throughout their interior than do non-magnetic models. This is not unexpected 
if we consider that the magnetic field acts to inflate the radius of the 
model. However, the reduction in mass density throughout the stellar interior
affects the sound speed profile, which has implications for seismic analyses. 

\begin{figure}[t]
    \plotone{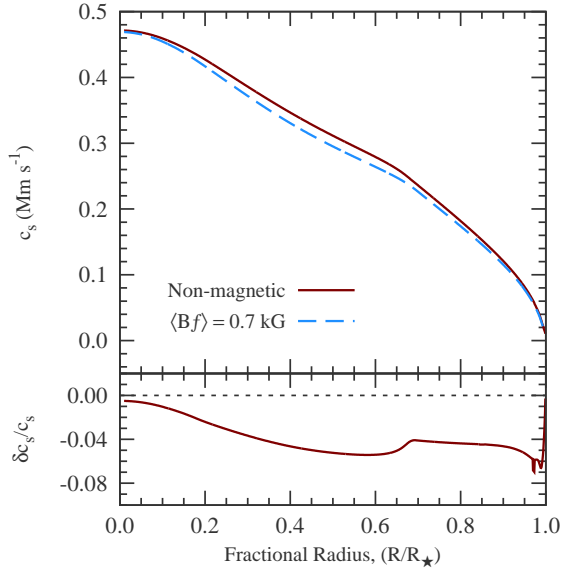}
    \caption{Top: sound speed profile for a 
        $M=0.599\msun$ model with and without the presence of a magnetic 
        field. Bottom: relative difference in the sound speed
        profile between the two models with respect to the non-magnetic
        model. The magnetic model was computed with a surface magnetic 
        field strength strong enough to reconcile the model radius with 
        the observed radius of YY Gem ($0.7$ kG). 
        \vspace{0.5\baselineskip} \\        
        (A color version of this figure is available in the online journal.)
        }
    \label{fig:dcs_rad}
\end{figure}

We plot the sound speed profile 
for two $M=0.6\msun$ models in Figure \ref{fig:dcs_rad}. The sound speed
is defined as
\begin{equation}
    c_{s}^2 = 
     \frac{P}{\rho}\left(\frac{\partial\ln P}{\partial\ln\rho}\right)_{\rm ad}
\end{equation}
where $P$ is the total pressure (gas + radiation + magnetic) and $\rho$ 
is the mass density. Using the total pressure to define the sound speed
will suffice for exploratory purposes.
The models have
a scaled-solar composition of [Fe/H] = $-0.2$ dex. One model has a magnetic
field with a surface field strength of $0.7$ kG using the turbulent dynamo
formulation. Note that the density profile for these two models are shown
in Figure \ref{fig:int_prof}.

Comparing the sound speed between the two models, we see that the magnetic 
model has a sound speed that is slower by about 5\%. The bottom panel in
Figure \ref{fig:dcs_rad} shows the relative sound speed difference between
the two models with respect to the non-magnetic models. We define
\begin{equation}
 \frac{\delta c_s}{c_s} = \frac{c_{s,\, {\rm mag}} - c_{s,\, 0}}{c_{s,\, 0}},
\end{equation}
where $c_{s,\, 0}$ is the sound speed in the non-magnetic model and 
$c_{s,\, {\rm mag}}$ is that for the magnetic model. The change in sound
speed between the two models affects the $p$-mode frequencies and frequency 
spacing, which consequently alters the interpretation of seismic data.

Imagine that observational data is obtained for a $M=0.6\msun$ star
that has a $0.7$ kG magnetic field. These two facts about the star are 
unknown to the observer. Instead, the observer has taken great care to 
obtain high resolution spectra with sufficiently high signal-to-noise.
From this data they are able to ascertain the effective temperature, 
$\teff$, and the approximate composition. They also have precise 
photometric data that allows for a seismic analysis. To first order, they 
are able to obtain the stellar radius from the large frequency splittings. 
The properties they derive for the star are $R = 0.619\pm0.006\rsun$, 
$\teff = 3\, 820 \pm 100$ K, and [Fe/H] = $-0.2\pm0.1$.\footnote{
Conveniently, these are the properties of YY Gem.}

With these properties, what inferences would the observer make about the star?
Using a grid of non-magnetic Dartmouth models \citep{Dotter2008,FC12}  
the observer would estimate the mass of the star to be roughly $M = 0.67\pm0.02\msun$. 
Or approximately 10\% larger than the actual mass. The age of the star would 
be difficult to estimate in this particular case due to the 
lack of significant radius evolution of low-mass stars on the MS.
One concession, is that the models would not match the observed $\teff$ 
within the $1\sigma$ limit quoted above. However, one can easily find a 
match within $2\sigma$. Moreover, the sound speed profiles between the 
higher mass non-magnetic models and the lower mass magnetic models are 
quite similar.

As we have seen in Section \ref{sec:yygem}, given a radius, $\teff$, and 
metallicity, one should be able to derive the appropriate mass for this 
star. With magnetic models, a slight degeneracy in mass and magnetic field 
strength would be introduced, as varying the magnetic field at a given mass 
can alter the radius and $\teff$. However, the true solution would be encompassed
by the formal uncertainties. Regardless, it will be more accurate than if
only non-magnetic models were used. 

Future investigations exploring the effects of a magnetic perturbation on
the pulsation analysis and the use of magnetic models for mass determinations 
are warranted. While the case above was exploratory and does not
represent the typical stars selected for asteroseismic analysis, the 
conclusions should be valid for all stars with an outer convective envelope.
Magnetically active stars are also not favored for asteroseismology. There
are issues with removing possible photometric variations due to spots. However,
the lack of photometric variability from spots is not necessarily indicative
of an inactive star \citep{Jackson2013}. 

\subsection{Exoplanet Properties}

The primary effect of introducing a magnetic perturbation
is that model radii are increased and $\teff$s are decreased. What has 
not been discussed is that model luminosities show a marginal decrease. 
This means the radius increase does not precisely balance the cooling of
the $\teff$, in contrast with what has been assumed previously 
\citep[see, e.g.,][]{Morales2008}. While there is no luminosity change 
immediately after the perturbation is introduced, the reduction in luminosity
occurs over thermal timescales. The models readjust their internal structure 
to compensate for the reduction of convective heat flux. These changes in 
the stellar surface properties can influence studies of extra-solar 
planets (exoplanets) around low-mass stars.

Transiting exoplanets have radii that are directly proportional to the 
radii of their host stars \citep[e.g.,][]{Seager2003}. Determining the
radius of a low-mass host star is difficult. Interferometry allows for 
the direct measurement of the stellar radius, but is only available for
the brightest (i.e., nearest) targets \citep{vonBraun2012,Boyajian2012}. 
One could attempt to perform an asteroseismic analysis, but it would suffer 
due to the intrinsic faintness of low-mass stars and also from the considerable 
photometric variability that is characteristic of these stars. In lieu of 
other methods, stellar evolution models are often used to aid in the interpretation 
of transiting exoplanetary systems \citep[e.g.,][]{Muirhead2012,Gaidos2013,
Dressing2013}.

Owing to a lack of available magnetic model grids, these studies have relied
exclusively on standard stellar evolution models. The use of magnetic models
can have an impact on these studies. For instance, \citet{
Muirhead2012} measured the metallicity and $\teff$ for the low-mass stars
tagged as planet candidates in the \emph{Kepler} mission \citep{Batalha2013}.
They then interpolated within Dartmouth isochrones \citep{Dotter2008} to 
determine stellar radii.
Had they used magnetic stellar models, for a given $\teff$, the stellar
radius would be larger, with the precise factor dependent on the magnetic
field strength assumed. This would result in larger, less dense (and 
therefore less Earth-like) planets. 

Estimating the surface magnetic field is the largest hindrance to using
magnetic models. Stars in the \emph{Kepler} field are typically distant,
meaning X-ray measurements will not necessarily be available. Deep observations
of the \emph{Kepler} field with \emph{Chandra} or \emph{XMM-Newton} could 
provide X-ray magnetic activity information. Alternatively, it might be 
feasible to obtain optical spectra and use the H$\alpha$ equivalent widths 
to estimate magnetic activity. H$\alpha$ is less revealing, as we are not 
aware of a direct relation linking H$\alpha$ equivalent widths to magnetic 
field strengths. However, it has been shown that H$\alpha$ equivalent widths 
correlate, to some extent, with X-ray luminosity \citep{Delfosse1998,
Reiners2007,Stassun2012}. Regardless, the presence of H$\alpha$ in emission 
would be indicative of an active star. At that point, several magnetic 
models could be used to constrain the stellar radius.

\section{Conclusions}
\label{sec:summ}

This paper addressed the question of how the presence of a magnetic
field affects the structure of low-mass stars with a radiative core. We 
approached this problem by taking a careful look at three DEB systems that 
show significant radius and $\teff$ deviations from standard stellar 
evolution models. Using the magnetic stellar evolution models introduced 
in a previous paper \citep{FC12b}, we attempted to reproduce 
the observational properties of UV Psc, YY Gem, and CU Cnc.

After finding that the magnetic models were able to reconcile the stellar
models with the observed radii and $\teff$s \citep[consistent with the
findings of][]{MM01}, it was shown in Section \ref{sec:surf_field_str} that 
the surface magnetic field strengths were likely too strong. This was determined 
by taking the coronal X-ray luminosity as an indirect diagnostic for the 
surface magnetic field strength \citep{Fisher1998,Pevtsov2003}. In contrast 
to the surface magnetic fields, the interior field strengths are of a 
plausible magnitude, consistent with the range of field strengths predicted 
to be within the Sun by 3D MHD models \citep{Brandenburg2005}.

We then attempted to reduce the surface magnetic field strengths while 
still maintaining the newfound agreement between stellar models and DEB
observations. The most plausible explanation we uncovered is that kinetic
energy in turbulent convective flows is actively converted into magnetic 
energy. Introducing changes to MLT that mimic this physical 
process yielded accurate models with surface magnetic field strengths nearly 
equal to those predicted by X-ray emission. We were unable, however, to 
implement this prescription for the stars of UV Psc. This is likely a 
consequence of the chosen magnetic field profile, which caused the models
to fail to converge. We note that by invoking the ``turbulent dynamo'' 
approach, we are able to generate substantial radius inflation with relatively
modest (sub-kG) magnetic fields. 

Beyond direct comparisons to observational data, we explored additional 
implications of the present study.  We found that different theoretical 
descriptions of the physical manifestation of magneto-convection lead to 
similarly inflated stars  and produce nearly identical stellar interiors. 
This was evidence by the radial density profile for models invoking the
methods presented in \citet{FC12b}, Section \ref{sec:des} (this work), and 
\citet{Chabrier2007}, which were all nearly identical. The latter uses a reduced 
mixing length approach to simulate the introduction of a turbulent 
dynamo. We provide a variable, $\Lambda$, to translate a reduction in mixing
length into a magnetic field strength (normalized to the equipartition 
field strength, which depends on the convective velocity).

We also argue that asteroseismic studies should be relatively unaffected 
by structural changes in the stellar models. Asteroseismic studies typically 
know the stellar composition and effective temperature. Anomalous values 
of the $\teff$ due to magnetic fields should provide a clue that magnetic 
fields are required. For a given density profile, it is difficult to reproduce 
the $\teff$ for a star of a given mass without a magnetic field. If this 
happens, it might be a hint to adopt magnetic models. Otherwise, if the 
$\teff$ is ignored, then masses would typically be overestimated by about 
10\%. 

Magnetic models will, however, have an immediate impact on the estimated 
radii of transiting exoplanets. The radii of transiting exoplanetary radii 
scale with the stellar radius. This is but one reason that further exploration 
of the role of magnetic fields in low-mass stars is required. Interpretation 
of low-mass star observations depend critically on the accuracy of stellar
evolution model predictions.  

\acknowledgements
G.A.F. thanks A. Brandenburg for the discussions that lead to the development
of the turbulent dynamo models advanced in this paper. The authors thank
the anonymous referee whose comments, questions, and suggestions improved 
the manuscript.
The authors also thank the William H. Neukom 1964 Institute for Computational 
Science at Dartmouth College and the National Science Foundation (NSF) 
grant AST-0908345 for their support. This research has made use 
of NASA's Astrophysics Data System, the SIMBAD database, operated at CDS, 
Strasbourg, France, and the \emph{ROSAT} data archive tools hosted by the 
High Energy Astrophysics Science Archive Research Center (HEASARC) at 
NASA's Goddard Space Flight Center.

\appendix
\section{Magnetic Flux Scaling Relation}
\label{app:phi_lx}

\begin{deluxetable*}{l l l c c c c c c c c}[b]
    \tablewidth{2\columnwidth}
    \tablecaption{Low-mass Stars from Reiners (2012) with Direct Magnetic Field
      Measurements used to Derive our $\Phi-L_x$ Relation.}
    \tablehead{
        \colhead{Star} & \colhead{Other} & \colhead{SpT} & \colhead{$\langle Bf \rangle$} & \colhead{$R_\star$} & \colhead{$\log(\Phi)$} & \colhead{$\pi$} & \colhead{$X_{\rm cr}$} & \colhead{HR} & \colhead{$\log(F_x)$} & \colhead{$\log(L_x)$} \\
        \colhead{Name} & \colhead{Name} & & \colhead{(kG)} & $(\rsun)$ & (Mx) & (mas) & $\mathrm{(cnts\, s^{-1})}$ &  & $\mathrm{(erg\, s^{-1}\, cm^{-2})}$ & $\mathrm{(erg\, s^{-1})}$
    }
    \startdata
HD 115383    & 	59 Vir & G0	  &  0.5	& 0.74 &	25.22 &	56.95  & 1.12 &	-0.14 &	-11.07 &	29.50 \\
HD 115617    & 	61 Vir & G6	  &  0.1	& 0.95 &	24.74 &	116.89 & 0.01 &	-0.96 &	-13.34 &	26.60 \\
$\sigma$ Dra & 	-	   & K0	  &  0.1	& 0.80 &	24.59 &	173.77 & 0.26 &	-0.80 &	-11.97 &	27.62 \\
40 Eri	     &  -	   & K1	  &  0.1	& 0.75 &	24.53 &	200.62 & 0.80 & -0.28 &	-11.26 &	28.21 \\
$\epsilon$ Eri&  -	   & K2	  &  0.1	& 0.70 &	24.59 &	310.94 & 2.82 &	-0.44 &	-10.77 &	28.32 \\
LQ Hya	     &  -	   & K2	  &  2.5	& 0.70 &	25.86 &	53.70  & 2.73 &	-0.04 &	-10.66 &	29.96 \\
GJ 566 B     &$\xi$ Boo B& K4 &  0.5	& 0.70 &	25.14 &	149.26 & 2.44 &	-0.31 &	-10.79 &	28.94 \\
Gl 171.2 A   & 	-	   & K5	  &  1.4	& 0.65 &	25.56 &	55.66  & 2.69 &	-0.04 &	-10.66 &	29.93 \\
Gl 182	     &  -      & M0.0 &  2.5	& 0.60 &	25.74 &	38.64  & 0.65 &	-0.19 &	-11.32 &	29.58 \\
Gl 803	     &  AU Mic & M1.0 &  2.3	& 0.50 &	25.54 &	100.91 & 5.95 &	-0.07 &	-10.33 &	29.74 \\
Gl 569 A	 &  -	   & M2.0 &  1.8	& 0.40 &	25.24 &	-	   & 0.49 &	-0.40 &	-11.52 &      -   \\
Gl 494	     &  DT Vir & M2.0 &  1.5	& 0.40 &	25.16 &	85.54  & 1.57 &	-0.01 &	-10.89 &	29.33 \\
Gl 70	     &  -	   & M2.0 &  0.2	& 0.33 &	24.12 &	87.62  & 0.04 &	-0.67 &	-12.68 &	27.51 \\
Gl 873	     &  EV Lac & M3.5 &  3.8	& 0.31 &	25.35 &	195.22 & 5.83 &	-0.16 &	-10.36 &	29.14 \\
Gl 729	     &V1216 Sgr& M3.5 &  2.1	& 0.20 &	24.71 &	336.72 & 0.94 &	-0.43 &	-11.25 &	27.78 \\
Gl 87	     &  -	   & M3.5 &  3.9	& 0.30 &	25.33 &	96.02  & -    &   -	  &   -	   &      -   \\
Gl 388	     &  AD Leo & M3.5 &  3.0	& 0.39 &	25.44 &	213.00 & 3.70 &	-0.27 &	-10.59 &	28.83 \\
GJ 3379	     &  -	   & M3.5 &  2.3	& 0.25 &	24.94 &	190.93 & 0.40 &	-0.20 &	-11.54 &	27.98 \\
GJ 2069 B    & 	CV Cnc & M4.0 &  2.7	& 0.25 &	25.01 &	78.10  & 0.24 &	-0.14 &	-11.74 &	27.95 \\
Gl 876	     &  IL Aqr & M4.0 &  0.2	& 0.31 &	24.07 &	213.28 & -	  &   -	  &   -	   &      -   \\
GJ 1005 A    & 	-	   & M4.0 &  0.2	& 0.23 &	23.81 &	191.86 & -	  &   -	  &   -	   &      -   \\
Gl 490 B	 &G 164-31 & M4.0 &  3.2	& 0.20 &	24.89 &	50.00  & 0.84 &	-0.22 &	-11.22 &	29.46 \\
Gl 493.1	 &  FN Vir & M4.5 &  2.1	& 0.20 &	24.71 &	123.10 & 0.14 &	-0.16 &	-11.98 &	27.92 \\
GJ 4053	     &LHS 3376 & M4.5 &  2.0	& 0.17 &	24.55 &	137.30 & 0.06 &	-0.49 &	-12.46 &	27.34 \\
GJ 299	     &  -	   & M4.5 &  0.5	& 0.18 &	23.99 &	148.00 & -	  &   -	  &   -	   &      -   \\
GJ 1227	     &  -	   & M4.5 &  0.2	& 0.19 &	23.64 &	120.00 & -	  &   -	  &   -	   &      -   \\
GJ 1224	     &  -	   & M4.5 &  2.7	& 0.18 &	24.73 &	132.60 & 0.21 &	-0.45 &	-11.91 &	27.93 \\
Gl 285	     &  YZ Cmi & M4.5 &  4.5	& 0.30 &	25.39 &	167.88 & 1.47 &	-0.21 &	-10.97 &	28.65 \\
GJ 1154 A    & 	-	   & M5.0 &  2.1	& 0.20 &	24.71 &	-	   & 0.10 &	-0.23 &	-12.15 &      -   \\
GJ 1156	     &  GL Vir & M5.0 &  2.1	& 0.16 &	24.51 &	152.90 & 0.13 &	-0.25 &	-12.04 &	27.67 \\
Gl 905	     &  HH And & M5.5 &  0.1	& 0.17 &	23.24 &	316.70 & 0.18 &	0.15  & -11.79 &	27.29 \\
GJ 1057	     &  CD Cet & M5.5 &  0.1	& 0.18 &	23.29 &	117.10 & -	  &   -	  &   -	   &      -   \\
GJ 1245 B    & 	-	   & M5.5 &  1.7	& 0.14 &	24.31 &	220.00 & 0.20 &	-0.37 &	-11.90 &	27.50 \\
GJ 1286	     &  -	   & M5.5 &  0.4	& 0.14 &	23.68 &	138.30 & -	  &   -	  &   -	   &      -   \\
GJ 1002	     &  -	   & M5.5 &  0.2	& 0.13 &	23.31 &	213.00 & -	  &   -	  &   -	   &      -   \\
Gl 406	     & CN Leo  & M5.5 &  2.4	& 0.13 &	24.39 &	418.30 & 0.23 &	-0.22 &	-11.79 &	27.05 \\
Gl 412 B	 &  WX Uma & M6.0 &  3.9	& 0.13 &	24.60 &	206.94 & 0.18 &	-0.64 &	-12.05 &	27.39 \\
GJ 1111	     &  DX Cnc & M6.0 &  1.7	& 0.12 &	24.17 &	275.80 & -	  &   -	  &   -	   &      -   \\
Gl 644C	     &  VB 8   & M7.0 &  2.3	& 0.10 &	24.15 &	153.96 & -	  &   -	  &   -	   &      -   \\
GJ 3877	     &LHS 3003 & M7.0 &  1.5	& 0.10 &	23.96 &	157.80 & -	  &   -	  &   -	   &      -   \\
GJ 3622	     &  -	   & M7.0 &  0.6	& 0.10 &	23.56 &	221.00 & -	  &   -	  &   -	   &      -   \\
LHS 2645	 &  -      & M7.5 &  2.1	& 0.08 &	23.91 &	-	   & -	  &   -	  &   -	   &      -   \\
LP 412-31    & 	-	   & M8.0 &  3.9	& 0.08 &	24.18 &	-	   & -	  &   -	  &   -	   &      -   \\
VB 10	     &V1298 Aql& M8.0 &  1.3	& 0.08 &	23.70 &	164.30 & -	  &   -	  &   -	   &      -   \\
LHS 2924	 &  -	   & M9.0 &  1.6	& 0.08 &	23.79 &	90.00  & -	  &   -	  &   -	   &      -   \\
LHS 2065	 &  -	   & M9.0 &  3.9	& 0.08 &	24.18 &	116.80 & -	  &   -	  &   -	   &      - 
    \enddata
    \label{tab:lms_xray_data}
\end{deluxetable*}

We compute the approximate unsigned surface magnetic flux for the dwarf stars 
presented in the \citet{Reiners2012a} review. The data are compiled in 
Table \ref{tab:lms_xray_data}. \citet{Reiners2012a} 
collected all of the reliably determined magnetic field measurements for cool 
dwarf stars. We select for our sample only those stars that had their
average surface magnetic field, $\langle Bf\rangle$, measured using 
Stokes $I$ polarization observations. We avoided using stars with estimates
provided from Stokes $V$ observations because Stokes $I$ 
yields a more accurate estimate of $\langle Bf\rangle$ \citep{Reiners2009}.
Since our models predict $\langle Bf\rangle$, Stokes $I$ observations
give a more direct comparison. 

We cross-correlated objects from \citet{Reiners2012a} 
with the {\it ROSAT} Bright and Faint Source Catalogues \citep{Voges1999,Voges2000} 
to extract X-ray count rates ($X_{\rm cr}$) and HRs. 
Objects that did not have an X-ray counterpart were excluded from our final
sample. We identified
all of the stars in the remaining subset that had parallax estimates from
\emph{Hipparcos} \citep{hip07}. Our final sample consists of 25 objects
with both X-ray counterparts and parallax measurements. Two additional 
data points for G-dwarfs, not found in \citet{Reiners2012a}, were 
included \citep{Anderson2010} bringing the total number of objects
to 27. Conversion from raw $X_{\rm cr}$, HR, and parallaxes to X-ray luminosities
is described in Section \ref{sec:surf_field_str}. The full sample of objects, 
their magnetic field properties, and
X-ray properties are given in Table \ref{tab:lms_xray_data}.

Results of extending the scaling relation with our sample are presented 
in Figure \ref{fig:lx_mx_rad}. The data used by \citet{Pevtsov2003} is 
shown for reference. Our data follows the same trend, but is more populated
at the low $L_x$ end. We caution that the data presented in Figure 
\ref{fig:lx_mx_rad} suffer from uncertainty in the adopted stellar radii.
When possible, we adopted the radii quoted by the original sources, although
their procedure for assigning radii was not always clear. For stars
where the original references did not have quoted radii, we assigned
radii based on spectral type using interferometric data as a guide \citep{
Boyajian2012}. Using spectral types as an indicator for stellar properties
is hazardous, especially for M-dwarfs, but we assumed large radius uncertainties
that would likely encompass the true value. 

The data are also limited by the spatial resolution of {\it ROSAT}. As an example, 
CV Cnc is an mid-M-dwarf binary included in our final data set. Estimates 
of its X-ray luminosity include contributions from CU Cnc, as discussed 
in Section \ref{sec:cucnc1}. Only for CV Cnc did we correct for
this confusion. Overall, uncertainty 
in the stellar radius affects the magnetic flux while the lack of spatial 
resolution affects the X-ray luminosity. Therefore, we believe our 
estimates are robust, particularly given that Figure \ref{fig:lx_mx_rad} 
is plotted using logarithmic units, where a factor of two does not 
contribute to a significant shift in the data. 

The regression line and associated uncertainty limits shown in Figure 
\ref{fig:lx_mx_rad} were obtained by performing an ordinary least squares 
(OLS) regression. A standard OLS was selected as we are 
not attempting to establish a causal relationship between $\Phi$ and $L_x$
\citep{Isobe1990}. Our primary concern is developing a relationship that 
permits a prediction of $\Phi$ given a value for $L_x$. Following the
recommendations of \citet{Isobe1990} and \citet{Feigelson1992}, we 
perform an OLS$\left(\log_{10}\Phi | \log_{10} L_x\right)$, where 
$\log_{10} L_x$ is the predictor variable and $\log_{10}\Phi$ is the
variable to be predicted. Note that we transformed the variables to a 
logarithmic scale because the data extends over several orders of magnitude 
in $\Phi$ and $L_x$.

The result of the OLS analysis yielded a regression line 
\begin{align}
    \log_{10}\Phi = & \left(24.873\pm0.004\right) + \left(0.459\pm0.018\right) \nonumber \\
         & \times \left(\log_{10}L_{x} - \mu_x\right),
\end{align}
where $\mu_x$ is the mean value of $\log_{10}L_{x}$ taken over the entire 
sample. A shift of the dependent variable was 
performed prior to the regression analysis. By doing this, we were able to 
minimize the error associated with the $y$-intercept. However, the 
standard deviation of the mean, $\sigma_x$, becomes the largest source
of error. We found $\mu_x = 28.34\pm0.97$, where we took
\begin{equation}
    \sigma_x = \sqrt{\frac{1}{N-1}\displaystyle\sum_{i = 1}^{N} (x_i - \mu_x)^2}.
\end{equation}
We opted for computing the standard deviation with $N - 1$ because of our
small sample (27 data points) so as to provide a conservative error estimate. \\

\twocolumngrid

\end{document}